\documentclass[12pt]{elsart} 
 
\usepackage{psfig} 
 
\begin{document}

\begin{frontmatter} 
 
\title{Comparison of Form Factors Calculated with Different Expressions 
for the Boost Transformation } 
\author[Amghar]{A. Amghar},  
\author[Desplanques]{B. Desplanques} 
\ead{desplanq@isn.in2p3.fr}, 
\author[Theussl]{L. Theu{\ss}l} 
\ead{Lukas.Theussl@uv.es}  
\address[Amghar]{Facult\'e des Sciences, Universit\'e de Boumerdes, \\ 
35000 Boumerdes, Algeria} 
\address[Desplanques]{Institut des Sciences Nucl\'eaires (UMR CNRS/IN2P3--UJF),  
\\  
F-38026 Grenoble Cedex, France} 
\address[Theussl]{Departamento de Fisica Teorica, Universidad de Valencia  \\  
E-46100 Burjassot (Valencia), Spain}

\begin{abstract} 
The effect of different boost expressions is considered for the calculation 
of  the ground-state form factor of a two-body system made of scalar  
particles interacting via the exchange of a scalar boson. The aim is to 
provide  an uncertainty range on  methods employed in implementing these 
effects as well  as an insight on their relevance when an ``exact'' 
calculation is possible.  Using a wave function corresponding to 
a mass operator that has the appropriate properties to construct 
the generators of the Poincar\'{e} algebra in the framework of 
relativistic quantum mechanics, form factors are calculated using 
the boost  transformations pertinent to the instant, 
front and point forms of this approach. Moderately and  strongly 
bound systems are considered with masses of the exchanged boson taken 
as zero, 0.15 times  the constituent mass $m$, and infinity. In the first 
and last cases, a  comparison with ``exact'' calculations is made 
(Wick-Cutkosky model and Feynman  triangle diagram). Results with a 
Galilean boost are also given.  Momentum transfers up to $Q^2=100\,m^2$ 
are considered. Emphasis is put on the contribution of the single-particle 
current, as usually done. It is  found that the present point-form 
calculations of form factors  strongly deviate from all the other ones, 
requiring large contributions from two-body currents.  
Different implementations of the point-form approach, where the role of these 
two-body currents would be less important, are sketched. 
\end{abstract} 
 
\end{frontmatter} 
 
\section{Introduction} 
Implementing relativity in the description of form factors of few-body  
systems is an important task nowadays. Motivated for some part by  
experiments performed at high $Q^2$ at different facilities, especially  
JLab, this ingredient is  required for correctly analyzing the  
measurements. Many approaches have been proposed, ranging  
from field-theory based to relativistic-quantum mechanics  
ones~\cite{Zuilhof:1980ae,Gross:1983yt,Karmanov:1992fv,Carbonell:1998rj,Licht:1970pe,Friar:1973,Chung:1988my,Chung:1988mu,Lev:1995,Klink:1998},  
implying sometimes seemingly founded approximations.  
A full realistic calculation is difficult and since the calculation of the  
deuteron form factors by Tjon and Zuilhof~\cite{Zuilhof:1980ae}, 
there are not so many  calculations that have reached the same level 
of correctness and,  
simultaneously,  of complexity. Full calculations can also be performed in  
other cases~\cite{Desplanques:2001zw,Desplanques:2001ze}, employing 
solutions of the  
Bethe-Salpeter equation~\cite{Salpeter:1951sz} with a zero-mass boson  
(Wick-Cutkosky model~\cite{Wick:1954eu,Cutkosky:1954ru}) or an infinite one  
(Feynman triangle diagram). They are  
somewhat academic but turn out to be quite useful as a testing ground of  
the approximate treatment of relativity in many studies. 
 
There are currently many recipes aiming to implement relativity in the  
description of some system with a reduced amount of effort. They imply 
for  instance minimal relativity factors, $\sqrt{m/e}$, or relativistic 
energies in  place of non-relativistic ones. When considering elastic 
form factors, that are  of main interest here, one also has to deal 
with boosting appropriately the  final state with respect to the 
initial one. Here too, one has looked for  simple  recipes. In the 
case of spin-less particles, the first effect one thinks of is  
the Lorentz-contraction. To account for it, it was proposed to 
make the  following replacement in the non-relativistic 
calculation~\cite{Licht:1970pe}
\begin{equation}  
F^{NR}(Q^2)\rightarrow F^{Rel}(Q^2)= F^{NR}(Q^2/[1+Q^2/(4M^2)]), 
\label{11a} 
\end{equation}  
where M is the total mass of the system under consideration.  
 
This relation, that is still being used 
\cite{Acus:2000ah,Santopinto:2002}, can be recovered 
somewhat easily (it is obtained by a simple  change of variable). 
Perhaps for this reason, it was  believed to represent a realistic 
way to implement an important feature of relativity 
(see some  discussion in Ref.~\cite{Friar:1973}). However,  it was 
realized later  on that  this prescription leads to constant  form 
factors at high $Q^2$, in disagreement with what is expected from the  
consideration of the Born amplitude. This one underlies predictions 
of power law behavior of form factors at large $Q^2$ ($Q^{-2}$ 
and $Q^{-4}$ for the pion  and nucleon form factors in QCD, 
respectively~\cite{Lepage:1979zb,Lepage:1979za},  or $Q^{-4}$ for the  
ground state of a system of scalar particles exchanging a scalar boson, 
as  considered in this work~\cite{Alabiso:1974sg}). On the other hand, 
Gl{\"o}ckle and Hamme~\cite{Hamme:1992} (see also~\cite{Glockle:1987hb}) 
analyzed solutions obtained from some Hamiltonian for  a similar system. 
They found support for the Lorentz contraction but could not 
conclude whether the deviations were due to the approximate nature 
of their Hamiltonian.  
 
Quite recently, another way to implement relativity in the calculation  
of form factors was proposed by Klink~\cite{Klink:1998}. Supposed 
to be  based on the point-form approach, it essentially relies on  
kinematical boost transformations. It has been applied for calculating  
form factors of the  deuteron~\cite{Allen:2000ge},  
the nucleon~\cite{Wagenbrunn:2000es},  a two-body system composed of 
scalar particles exchanging a zero-mass boson~\cite{Desplanques:2001zw} 
and a system corresponding to a zero-range  
interaction~\cite{Desplanques:2001ze}. In the first case,  
there is no significant improvement with  
respect to  a non-relativistic calculation. In the second case, the  
agreement is at first sight spectacular, especially at the lowest $Q^2$  
range considered by the authors. However, one can guess a tendency  
to an increasing under-prediction of the measurements in the highest  
range. In the third and fourth cases, there is no experiment but  
comparison is possible  with what can be considered as an ``exact''  
calculation as far as relativity is concerned. Corrections originating  
from the field-theory character of the underlying model, such as  
constituent form factors~\cite{Cano:2001hy}, cancel out in the comparison  
while they should be accounted for in the nucleon case mentioned above.  
In the last two cases, an important discrepancy shows up, especially  
in the limit of a zero-mass system or at large $Q^2$, which a  
relativistic approach should deal with. Two features emerge from these  
form factors when compared to  ``exact'' calculations (as well as  
non-relativistic ones): point-form results evidence an increased  
charge radius and a rapid fall-off, faster than  
expected from the Born-amplitude.  
 
In the present paper, we want to see in particular whether the peculiar 
features evidenced by recent calculations of form factors are specific 
to the point-form approach or characterize also other forms of relativistic 
quantum mechanics, such as the instant- and front-form. In 
this order, one can start from some  mass operator, which enters in 
the construction of the generators of the Poincar\'{e} 
group~\cite{Bakamjian:1953kh,Keister:1991sb}.  However, 
this does not indicate the connection between the dynamical variables 
entering this construction and the experimentally physical 
quantities~\cite{Coester:1975hj,Sokolov:1985jv}. To establish this 
connection, a physical model has to be considered, while the underlying 
field-theory could provide an ``exact''calculation of form 
factors, somewhat playing the role of a measurement. Thus, the 
interaction entering the mass operator should fulfill two different 
constraints. Besides those required to construct the generators of the 
Poincar\'{e} group, we demand that the interaction allows 
one to recover the field-theory one in the small coupling limit. 
This is the least that we can ask.

Steps of the above program can be found in the 
literature~\cite{Bakamjian:1953kh,Keister:1991sb} (see also for 
instance refs.~\cite{Carbonell:1998rj,Wallace:2001nv} in a slightly 
different context). For our purpose, we 
give some of them here, both for comprehensiveness and definiteness. 
As we want to know how an approach different from the point-form one 
does in comparison to an ``exact'' calculation, we will specialize 
in this work  on the instant-form approach. While doing so, we give 
a particular attention to the change of variable which allows 
one to express the momenta of two particles, let's say $\vec{p}_1$ 
and $\vec{p}_2$ for a two-body system, in terms of the total momentum, 
$\vec{P}$, and an internal variable, $\vec{k}$, which enters the 
mass operator. This relation, represented by a unitary 
transformation~\cite{Coester:1975hj}, is nothing but the one 
introduced by Bakamjian and Thomas~\cite{Bakamjian:1953kh}. It is 
essential for determining how a wave function obtained from some 
mass operator transforms when the system under consideration is 
boosted. The condition that the mass spectrum be independent of 
the momentum of the system is found to provide the standard 
constraints on the mass operator, from which the Bakamjian-Thomas 
construction of the generators of the 
Poincar\'{e} group immediately follows~\cite{Bakamjian:1953kh}. 
The mass operator so obtained can then be used for  the front- 
or point-form approaches~\cite{Keister:1991sb}.

When available, we will give results of what can be considered 
an ``exact''  calculation (Wick-Cutkosky model, Feynman triangle 
diagram), the corresponding  physics  being supposed to be 
accounted for by the different approaches listed above.  Due to 
the difficulty to fully account for this physics in the finite-mass 
boson case (the  bound-state  
spectrum is not reproduced by the interaction model used to calculate  
the wave function employed in our study, and current conservation 
is not  necessarily fulfilled), we will also present results obtained  
with a tentatively improved interaction. The comparison of the results  
will allow one to get some insight into the uncertainty due to the  
approximate description of the interaction. This aspect is in any  
case difficult to assess in quantum mechanics approaches which 
have to account  effectively, in one way or another, for the 
field-theory character of the  original model. 

Concerns have been expressed about comparing approaches as different as 
field-theory and relativistic quantum mechanics, which correspond to 
problems with an indefinite and a fixed number of particles. We believe 
that this is what should be done if one wants to check the reliability 
of some relativistic quantum mechanics approach. When predictions of these 
approaches are compared to a measurement, they imply the underlying 
physics, QED, some meson theory or QCD. In all cases, this physics is 
field-theory based. We could also notice that comparisons similar to 
the present one have started long ago. Coester and Ostebee, for instance, 
noticed some agreement of their instant-form calculation for the 
deuteron form factor at low $Q^2$~\cite{Coester:1975hj} with a 
field-theory one by Gross~\cite{Gross:1966fg}. Finally, we stress
that the two approaches we are using are developed independently. 
There is no attempt to make some reduction of the Bethe-Salpeter 
equation, which is sometimes done. As noticed for instance by Lev (see 
introduction of ref.~\cite{Lev:1994vf}), the description of a 
bound system in relativistic quantum mechanics  represents an 
approximation to the one based on the Bethe-Salpeter equation 
but it offers the advantage that it preserves the covariance 
while the above reduction generally does not. 

When comparing form factors calculated from different relativistic quantum 
mechanics approaches, we have in mind that, ultimately, they should coincide, 
in agreement with the idea that these approaches  are equivalent, 
up to a unitary transformation \cite{Sokolov:1977im}. 
The first step concerns the mass operator, already mentioned, 
whose solutions can be used in any scheme, only with different 
transformation properties when going from the variables, 
$\vec{p}_1,\;\vec{p}_2$,  to 
$\vec{P}, \;\vec{k}$. The second step concerns the current 
which consists of a one-body part and, for a two-body system, 
a two-body one, which is interaction dependent. The contribution 
of this second part, when added to the former, should ensure the expected 
equivalence of the various relativistic quantum mechanics approaches. 
In this work, we will concentrate on the contribution of the 
single-particle current, most often retained in usual calculations 
of form factors. As significant differences appear at this level, 
making it difficult to realize the above unitary equivalence, 
we will give a particular attention to everything that can cast 
light on how these differences occur and manifest themselves. 

Concerning the observables, we will consider charge and  scalar form 
factors stemming from the coupling to  Lorentz-vector and Lorentz-scalar 
probes, respectively. We are still interested in the ground state of a 
system composed of scalar particles interacting via the exchange of a scalar 
boson. Results, obtained for instant- but also front-, and point-form 
approaches, will be compared to what could be considered as an ``exact'' 
calculation. As it is a current practice to compare relativistic 
calculations to non-relativistic ones, we also include in our study 
results obtained by applying a Galilean boost. The corresponding 
wave functions are obtained from an effective, theoretically-motivated 
interaction which does well for the  spectrum~\cite{Amghar:2000pc} 
and allows for a conserved current. The comparison  of the different 
results for charge and scalar form factors will  also evidence 
other features that are not related to the boost  itself but are 
more or less related to standard exchange currents.  These two-body 
currents, which are required,  among other things, to fulfill 
covariance, will be evoked at many places in the paper but will 
not be the object of a systematic study.  Some insight on their 
contribution will be obtained  by looking at the form factors 
calculated in a frame different from  the Breit one. 

As already mentioned, comparison of form factors obtained in 
the instant form of relativistic quantum mechanics, on which 
some emphasis is put here, and those obtained in a field-theory 
framework have already been done in the past~\cite{Coester:1975hj}.
The system under consideration in the present work will not be as 
realistic as the one looked at in this earlier work. Instead, we 
consider situations where relativity is expected to play an 
essential role such as large momentum transfers or strongly bound 
systems, partly in relation with Lorentz-contraction effects. 
Accordingly, we do not make any $1/M^2$ approximation in dealing 
with the boost transformation. 
 
The plan of the paper is as follows. In the second section, we 
concentrate on  the interaction and its reduction to an invariant mass 
operator. This interaction allows one to establish some relation with 
field-theory in the weak coupling limit and is used for the 
determination of wave functions required for the calculation 
of form factors. This is done within the instant-form approach 
while the relation to the work by Bakamjian and Thomas is 
emphasized. The third section is devoted to the  expression of the form 
factors we are calculating in different forms of relativistic quantum 
mechanics.  Results up to 
$Q^2=100\,m^2$ are presented in the fourth section. They concern  
the scalar and charge form factors of the ground state of the system  
under consideration.  Three masses of the exchanged boson, $\mu=0$  
(Wick-Cutkosky  model), $\mu=0.15\,m$ and $\mu=\infty$ are considered,
as well as  two masses for the system, $M\simeq 1.6\,m$, which implies 
a moderately bound system,  and $M=0.1\,m$, which is an extreme case 
that a  relativistic approach should in principle be able to deal with. 
The fifth section contains a discussion and  the conclusion. 
 
  
\section{Choice of the two-body interaction} 
The starting point to calculate form factors supposes to determine wave 
functions from some equation together with some interaction. Quite 
generally, the corresponding mass spectrum depends on the momentum of 
the system under consideration and, therefore, is not covariant. Using 
a Hamiltonian formalism with a one-boson exchange interaction, the 
dependence was studied in refs.~\cite{Hamme:1992,Glockle:1987hb}, in 
relation with Lorentz-contraction effects. Amazingly, the violation of 
covariance was found rather small, due probably to the choice of 
appropriate ingredients (relativistic energies, single-particle 
normalization factors $\sqrt{m/e}$, etc). This lack of covariance can 
be ascribed to an incomplete description of the interaction. It can be 
remedied, but this assumes to determine the corrections to the 
interaction order by order.

In the present work, we will proceed in a slightly different manner, 
where the above covariance problem is easily solved, although indirectly. 
Not surprisingly, one finds a close relation to the work by 
Bakamjian and Thomas~\cite{Bakamjian:1953kh} developed within the 
instant-form of relativistic quantum mechanics. There, 
the covariance is verified provided that the interaction fulfills 
some constraints that will be recovered. Once this interaction is 
obtained, the construction of the generators of the Poincar\'{e} 
group is straightforward. As this construction has been given in 
many places, it is not repeated here. For our instant-form 
calculations of form factors, the main point concerns the transformation 
properties of the wave function from one frame to another one. These ones  
are accounted for automatically by directly considering  the solutions 
of the appropriate equation in terms of the physical variables, rather 
than the variables most often introduced to construct the  generators 
of the Poincar\'{e} group and, especially, the mass operator. However, 
the consideration of this latter one may be useful for calculations of 
form factors in the front-form or point-form approaches~\cite{Keister:1991sb}. 
How to employ 
the solutions of the mass operator with this aim has been described  
in the literature~\cite{Chung:1988my,Chung:1988mu,Klink:1998}.

\subsection{Getting an invariant mass operator} 
 
Our starting point is an equation  with ingredients purposely  
chosen. It has the following form 
\begin{eqnarray} 
\nonumber  
\lefteqn{ 
\Big(E_P^2-(e_{p_1}+e_{p_2})^2\Big)\;\Phi(\vec{p}_1,\vec{p}_2)= 
\int\!\!\int \frac{d\vec{p}_1\!'}{(2\pi)^3}\;\frac{d\vec{p}_2\!'}{(2\pi)^3}}  
\\ && \qquad \qquad 
\frac{\sqrt{2\,(e_{p_1}+e_{p_2})}}{ \sqrt{2\,e_{p_1}} \, \sqrt{2\,e_{p_2}} }\, 
V_{int}(\vec{p}_1,\vec{p}_2,\vec{p}_1\!',\vec{p}_2\!')\; 
\frac{\sqrt{2\,(e_{p_1'}+e_{p_2'})}}{ \sqrt{2\,e_{p_1'}} \, \sqrt{2\,e_{p_2'}}  
}\; \Phi(\vec{p}_1\!',\vec{p}_2\!'), 
\label{20a} 
\end{eqnarray} 
where $E_P=\sqrt{M^2+\vec{P}^2}$, $e_p=\sqrt{m^2+\vec{p}\,^2}$. The 
quantities  $M$ and $\vec{P}$ represent the total mass and the total 
momentum of the system  under consideration. 

The quadratic dependence of the above equation 
on the energy or the phase-space  factors are not without any relation 
to other approaches. As will be seen below,  they  greatly facilitate the 
determination of the constraints that have to be  fulfilled in order 
to ensure the invariance of the mass spectrum while providing at 
the same time a direct way to relate a wave function corresponding to 
a finite  momentum, $\vec{P}$, to that one for the rest frame (c.m.).  
How to go back to the Hamiltonian formulation or to various mass 
operators is outlined at the end of the section. 

In the instant form, the momenta of the constituent particles are 
related to the total  momentum $\vec{P}$ by the equality: 
\begin{equation}  
\vec{p}_1+\vec{p}_2=\vec{p}_1\!'+\vec{p}_2\!'=\vec{P}, 
\label{20b} 
\end{equation} 
consistently with the property that the momentum has a kinematical 
character in this form. When the dynamics is described on a surface 
different from the instant-form one, $t=ct$, other relations between 
momenta are obtained. For a hyper-plane, $\lambda \cdot x=ct$ 
with $\lambda^2 =1$, for instance, the relation reads: 
\begin{equation}  
\vec{p}_1+\vec{p}_2-\frac{\vec{\lambda}}{\lambda^0}\;(e_1+e_2)= 
\vec{p}_1\!'+\vec{p}_2\!'-\frac{\vec{\lambda}}{\lambda^0}\;(e_1'+e_2')= 
\vec{P}-\frac{\vec{\lambda}}{\lambda^0}\;E_P. 
\label{20ba} \end{equation} 
The deviation from Eq.~(\ref{20b}) involves terms proportional 
to $E_P-e_1-e_2$ that, using Eq.~(\ref{20a}), can be turned into an 
interaction term, as expected from changing the hyper-surface on which 
physics is described. 
The square-root factors, $\sqrt{2\,e_{p_1}} 
\, \sqrt{2\,e_{p_2}}$, appearing at  the denominator in Eq.~(\ref{20a}) are 
well known normalizations for  single-particle states. Anticipating on the 
developments given below, the  factor, 
$\sqrt{2\,(e_{p_1}+e_{p_2})}$, is introduced to provide a total  
phase-space factor that is invariant under a boost. Provided  
$V_{int}(\vec{p}_1,\vec{p}_2,\vec{p}_1\!',\vec{p}_2\!')$ is appropriately 
chosen (see below),  
the above equation offers the great  
advantage that a change of variable allows one to  transform it into the  
center of mass:  
\begin{equation} 
(M^2-4\,e_{k}^2)\;\phi_0(\vec{k})= 
\int \frac{d\vec{k}'}{(2\pi)^3}\; 
\frac{1}{ \sqrt{e_{k}} }\; 
V_{int}(\vec{k},\vec{k}')\; 
\frac{1}{ \sqrt{e_{k'}} }\; 
\phi_0(\vec{k}'). 
\label{20c} 
\end{equation} 
The demonstration can be done in three steps concerning successively the energy  
factor at the l.h.s., the phase space factor at the r.h.s. and the 
interaction. Notice that the appearance  of the factors $1/\sqrt{e_{k}}$ at the  
r.h.s. of 
Eq.~(\ref{20c}) is fully consistent with unitarity in the continuum and 
a normalization of $V_{int}$ given by a standard boson propagator 
$1/(\mu^2-q^2)$.

\subsubsection{Energy factor} 
In describing a two-body system, one is used to  introduce the total and 
relative momenta defined as follows,  
$\vec{p}_1=\frac{1}{2} \vec{P}+\vec{p}$ and $\vec{p}_2=\frac{1}{2}  
\vec{P}-\vec{p}$. In terms of these variables, the energy factor 
appearing at the l.h.s. of Eq.~(\ref{20a})  can be written as: 
\begin{eqnarray} 
E_P^2-\left(e_{p_1}+e_{p_2}\right)^2 &=& E_P^2-4\,(m^2+p^2)-P^2+ 
\left(e_{\frac{1}{2}P+p}-  
e_{\frac{1}{2}P-p}\right)^2 \nonumber \\ &=&  
M^2-4\,(m^2+p^2)+\left(e_{\frac{1}{2}P+p}- e_{\frac{1}{2}P-p}\right)^2 
\label{20d}  
\end{eqnarray} 
where, here as well as throughout the paper, the notation, $e_{q\pm  
p}=\sqrt{m^2+  (\vec{q} \pm \vec{p})^2}$, is adopted for simplicity 
(arrows are omitted). 

As is seen, the r.h.s.  of the above expression depends on the 
total momentum of the system, $\vec{P}$, preventing one, a priori, 
from fulfilling covariance properties. This can be remedied however by 
introducing a different  change of variables, suggested by Bakamjian and 
Thomas~\cite{Bakamjian:1953kh}:
\begin{eqnarray}
\vec{p}_1&=&\vec{k}-\vec{P}\;\frac{\vec{k} \cdot \vec{P}}{P^2}
+\vec{P}\;\frac{\vec{k} \cdot \vec{P}}{P^2}\;\frac{\sqrt{4\,e_k^2+P^2}}{2\,e_k}
+e_k\;\frac{\vec{P}}{2\,e_k}, \nonumber \\
e_{p_1}&=&e_k\;\frac{\sqrt{4\,e_k^2+P^2}}{2\,e_k} 
+\vec{k} \cdot \frac{\vec{P}}{2\,e_k},
\label{20e}  
\end{eqnarray}
together with similar expressions for particle 2, where the 
change, $\vec{k} \rightarrow -\vec{k}$ has to be made.

The above transformation has the structure of a Lorentz transformation:
\begin{eqnarray}
\vec{p}_1&=&\vec{k}
+\vec{v}\;\frac{\vec{k} \cdot 
\vec{v}}{1-v^2}\;\frac{1}{\frac{1}{\sqrt{1-v^2}}+1}
+e_k\;\frac{\vec{v}}{\sqrt{1-v^2}}, \nonumber \\ 
e_{p_1}&=&e_k\;\frac{1}{\sqrt{1-v^2}} 
+\vec{k} \cdot \frac{\vec{v}}{\sqrt{1-v^2}},
\label{20e1}  
\end{eqnarray}
with an important difference however. The velocity dependent factor 
appearing in the transformation:
\begin{equation}
\frac{\vec{v}}{\sqrt{1-v^2}} =\frac{\vec{P}}{2\,e_k}, 
\label{20e2}  
\end{equation}
is determined by the c.m. kinetic energy of the constituents, $2\,e_k$, 
instead of the total mass of the system, $M$. As is seen from the equation:
\begin{equation}
\frac{\vec{v}}{\sqrt{1-v^2}}=\frac{\vec{P}}{M}
+\frac{\vec{P}}{M}\; \frac{M^2-4\,e_k^2}{2\,e_k\;(M+2\,e_k)}, 
\label{20e3}  
\end{equation}
the difference with a pure kinematical boost, like the one considered 
in a recent work by  Wallace~\cite{Wallace:2001nv}, is typically an 
interaction effect, see Eq.~(\ref{20c}). This one appears at one place 
or another, depending on the relativistic quantum mechanics approach.
A transformation similar to the above one, but  introduced in a 
slightly different context, can also be found in the literature  
(see for instance  ref.~\cite{Hamme:1992}).

Using the change of variables, Eq.~(\ref{20e}), it is easily checked 
that the energy factor can be cast into the desired form: 
\begin{equation} 
 E_P^2-(e_{p_1}+e_{p_2})^2= M^2-4\,e_k^2 \; . 
\label{20f}  
\end{equation} 
A related expression that may be useful is the following:  
\begin{equation} 
 \vec{k}^2= \left(\frac{\vec{p}_1-\vec{p}_2}{2}\right)^2-  
 \left(\frac{e_1-e_2}{2}\right)^2. 
\label{20fb}  
\end{equation} 
A similar expression is expected to hold in other forms of 
relativistic quantum  mechanics but with relations of the 
momenta $\vec{p}_1$ and $\vec{p}_2$ to the   total momentum $\vec{P}$ 
and the $\vec{k}$ variable different from those given  
in Eqs.~(\ref{20b},~\ref{20e}) (see Eqs.~(3.43, 3.44) in 
ref.~\cite{Carbonell:1998rj} for a  particular example in 
front-form dynamics).  It is also noticed that the variable $\vec{k}$, 
which is quite useful for separating the 
internal and external motions, cannot generally be identified with the 
relative momentum of a two-particle system. As noticed in ref. 
\cite{Coester:1975hj}, one should guard to attribute 
an ``observable significance to the variable so defined''. 
The remark may be relevant in understanding point-form results 
obtained in the present work. 
 
Coming back to the choice of the quadratic expression for the 
energy factor in Eq.~(\ref{20a}), we notice it is quite helpful 
because it  allows one to remove the term $\vec{P}^2$ appearing in 
both $E_P^2$ and  $(e_{p_1}+e_{p_2})^2$, leaving only terms 
of the form $\vec{P}\cdot\vec{p} $  that can be absorbed in the 
definition of the internal variable, $\vec{k}$. This would not 
be  the case with a linear dependence on $E_P$. 
 
\subsubsection{Phase-space factor} 
To deal with the phase space factors, it is appropriate to redefine the  
wave function,  
\begin{equation} 
\Phi(\vec{p}_1,\vec{p}_2)=  
\frac{  \sqrt{2\,(e_{\frac{1}{2}P+p}+ e_{\frac{1}{2}P-p}) }       }{  
\sqrt{ 2\,e_{\frac{1}{2}P+p} } \; \sqrt{2\,e_{\frac{1}{2}P-p} }    } \;  
\tilde{\phi}_P(\vec{p}\,). 
\label{20g} 
\end{equation} 
In this way, Eq.~(\ref{20a}) reads: 
\begin{eqnarray} 
\Big(E_P^2-(e_{\frac{1}{2}P+p}+ e_{\frac{1}{2}P-p})^2\Big) \;  
\tilde{\phi}_P(\vec{p}\,)= 
\nonumber \hspace{4cm} \\ 
\int \frac{d\vec{p}\,'}{(2\pi)^3}\; 
V_{int}(\vec{p},\vec{p}\,',\vec{P})\; 
\frac{  e_{\frac{1}{2}P+p'}+ e_{\frac{1}{2}P-p'}      }{  
 2\,e_{\frac{1}{2}P+p'}\; e_{\frac{1}{2}P-p'}     } \;  
\tilde{\phi}_P(\vec{p}\,'), 
\label{20h} 
\end{eqnarray} 
where the total and relative  momentum variables are used. The important point  
is that the phase-space factor can now be transformed into a simpler one,  
using the same variable transformation as for the energy factor discussed  
above, Eq.~(\ref{20e}): 
\begin{equation} 
\frac{d\vec{p}\,'}{(2\pi)^3}\;\frac{  e_{\frac{1}{2}P+p'}+ e_{\frac{1}{2}P-p'}  
    }{  
 2\,e_{\frac{1}{2}P+p'} \; e_{\frac{1}{2}P-p'}     }=  
\frac{d\vec{k}\,'}{(2\pi)^3}\; 
 \frac{1}{e_k}. 
\label{20i} 
\end{equation} 
Taking this relation into account, together with Eq.~(\ref{20f}),  
Eq.~(\ref{20h}) can be written:  
\begin{equation} 
\Big(M^2-4\,e_{k}^2\Big)\;\tilde{\phi}_P(\vec{p})= 
\int \frac{d\vec{k}'}{(2\pi)^3}\; 
V_{int}(\vec{p},\vec{p}\,',\vec{P})\; 
\frac{1}{ e_{k'} }\; 
\tilde{\phi}_P(\vec{p}\,'). 
\label{20j} 
\end{equation} 
To get an equation identical to Eq.~(\ref{20c}), it remains to demand 
that the interaction $V_{int}(\vec{p},\vec{p}\,',\vec{P})$ can be 
cast into the form  $V_{int}(\vec{k},\vec{k}\,')$. 
 
\subsubsection{Interaction} 
The simplest interaction that we can   think of is given by 
one-boson exchange  in the instantaneous approximation,  
which in the c.m. would read: 
\begin{equation} 
V=-g^2\, 
\frac{4\,m^2}{\mu^2+(\vec{k}-\vec{k}\,')^2} \; . 
\label{20m} 
\end{equation} 
As a starting point for our discussion,  we assume that the interaction in  
terms of the initial momenta, $\vec{p}$, is identified with the above  
one in the  
c.m.,  and is Lorentz-invariant in the case of free particles.  
This provides the following form: 
\begin{equation} 
V=-g^2\, 
\frac{4\,m^2}{\mu^2-\Big(\frac{1}{2}(p_1-p_2)-\frac{1}{2}(p'_1-p'_2)\Big)^2}  
\;. 
\label{20n} 
\end{equation} 
Consistently with an instant-form approach, this expression will be used 
with the relation between momenta given by Eq.~(\ref{20b}).  
 
Quite generally, the denominator in this equation differs from the one  
given  by the instantaneous interaction, Eq.~(\ref{20m}). The difference  
can be seen  when it is expressed as follows:  
\begin{eqnarray} 
\nonumber 
-\Big(\frac{1}{2}(p_1-p_2)-\frac{1}{2}(p'_1-p'_2)\Big)^2&=&   \\  
&& \hspace{-10em} 
(\vec{p}-\vec{p}\,')^2-\frac{1}{4} 
\Big( (e_{\frac{1}{2}P+p} - e_{\frac{1}{2}P-p}) 
   -(e_{\frac{1}{2}P+p'} - e_{\frac{1}{2}P-p'}) \Big)^2 
\nonumber  \\ &=&  
   \vec{k}^2+\vec{k}\,'^2- 2\,\vec{k}\cdot\vec{k}\,' + \Delta, 
\label{20o} 
\end{eqnarray} 
with  
\begin{eqnarray} 
   \Delta&=&    2 \, \frac{1}{4}\; 
   \left(e_{\frac{1}{2}P+p} - e_{\frac{1}{2}P-p}\right)\; 
   \left(e_{\frac{1}{2}P+p'} - e_{\frac{1}{2}P-p'}\right) 
   \nonumber \\& & \hspace{2cm} 
   -2\,\left(\frac{\vec{p}\cdot\vec{P}\;\vec{p}\,'\cdot\vec{P}}{\vec{P}^2} 
-\frac{\vec{k}\cdot\vec{P}\;\vec{k}\,'\cdot\vec{P}}{\vec{P}^2}\right). 
\label{20oo} 
\end{eqnarray} 
Noticing that:
\begin{equation}
\vec{p}\cdot\vec{P}=\vec{k}\cdot\vec{P}\;
\sqrt{\frac{4\,(m^2+\vec{k}^2)+\vec{P}^2}{4\,(m^2+\vec{k}^2)}}, 
\end{equation}
one can check that $\Delta$ is proportional to the 
square of the  total momentum, $\vec{P}$. By removing this term 
at the denominator of  the meson propagator in Eq.~(\ref{20n}), 
the resulting interaction can be transformed into an interaction  
depending only on the variables, $\vec{k}$ and $\vec{k},'$. 
In this way, Eq.~({\ref{20j}}) can now be written: 
\begin{equation} 
\Big(M^2-4\,e_{k}^2\Big)\;\tilde{\phi}_P(\vec{p}\,)= 
\int \frac{d\vec{k}'}{(2\pi)^3}\; 
V_{int}(\vec{k},\vec{k}')\; 
\frac{1}{ e_{k'} }\; 
\tilde{\phi}_P(\vec{p}\,'). 
\label{20ja} 
\end{equation} 
The equivalence with Eq.~(\ref{20c}) is achieved by putting: 
\begin{equation} 
\tilde{\phi}_P(\vec{p}\,)=\sqrt{ e_k } \; \phi_0(\vec{k}). 
\label{20k} 
\end{equation} 
This ensures that the mass spectrum of Eq.~(\ref{20a}) is independent 
of the  total momentum, $\vec{P}$,  and, at the same time, provides 
the relation between  the initial wave function and the one in the c.m.: 
\begin{equation} 
\Phi(\vec{p}_1,\vec{p}_2)= 
\sqrt{ \frac{ e_k\;(e_{\frac{1}{2}P+p} + e_{\frac{1}{2}P-p})}{ 2 \;  
e_{\frac{1}{2}P+p}  \;e_{\frac{1}{2}P-p}} } \; \;\phi_0(\vec{k})\; . 
\label{20l} 
\end{equation}  
While the present approach allows one to identify corrections 
that have to be  made in some interaction model to fulfill 
covariance, the properties of the resulting interaction, namely the 
dependence on the internal variables $\vec{k}^2$, $\vec{k'}\,^2$ and 
$\vec{k}\cdot\vec{k'}$ only,  are in complete  agreement 
with what was required on general grounds by Bakamjian and 
Thomas to construct the generators of the Poincar\'{e} 
group~\cite{Bakamjian:1953kh}. This construction then follows 
immediately.
 
\subsection{Relation to other equations} 
Some applications rather require a linearly energy-dependent equation. 
This can  be obtained from Eq.~(\ref{20a}) which, schematically, reads: 
\begin{equation} 
E_P^2\;\Phi(...) =\Big( (e_{p_1}+e_{p_2})^2 + V \Big)\;\Phi(...). 
\label{20q} 
\end{equation} 
It is straightforward to show that, for $E_P\geq 0$, $\Phi(...)$ is also  
solution of the following equation: 
\begin{eqnarray} 
E_P \; \Phi(...) & = &\sqrt{ (e_{p_1}+e_{p_2})^2 + V}\;\;\Phi(...)\nonumber \\ 
 & = &\left(e_{p_1}+e_{p_2}+  
\Big( \sqrt{ (e_{p_1}+e_{p_2})^2 + V}-e_{p_1}-e_{p_2}\Big)\right)\;\Phi(...), 
\label{20r} 
\end{eqnarray} 
where the last term, when it is expanded, has clearly an interaction 
character,  this one appearing at all orders as expected in the 
instant form of  relativistic quantum mechanics with a linearly 
energy dependent equation. 

We also notice that Eq.~(\ref{20c}), which the set of equations represented by 
Eq.~(\ref{20a}) reduces to, can be written as follows: 
\begin{equation} 
M^2\;\phi_0(\vec{k})= 4\,e_{k}^2\;\phi_0(\vec{k})+
\int \frac{d\vec{k}'}{(2\pi)^3}\; 
\frac{1}{ \sqrt{e_{k}} }\; 
V_{int}(\vec{k},\vec{k}')\; 
\frac{1}{ \sqrt{e_{k'}} }\; 
\phi_0(\vec{k}'). 
\label{20r2} 
\end{equation} 
In this form, the equation is identical to the one obtained 
with a mass  operator defined as $M^2=M_0^2+V'$, often referred 
to in the domain, with $M_0$ being the kinetic energy 
part~\cite{Keister:1991sb}. As above for the energy dependence, a 
linear version of this mass operator, $M=M_0+V''$, could be obtained.

\subsection{A few remarks} 
We notice that the last term in Eq.~(\ref{20o}), $\Delta$, defined by  
Eq.~(\ref{20oo}), is quite small in the non-relativistic limit (term of the  
order $1/m^6$). This can be better seen on the following alternative  
expression: 
\begin{equation} 
\Delta = -2 \; \frac{\vec{P} \cdot \vec{p} \;\; \vec{P} \cdot \vec{p}\,'  
\;(S-S' )^2}{S^2  \; S'^2  \; \Big(1-  \frac{P^2}{S \; S'\;} 
+ \sqrt{ 1-\frac{P^2}{S^2}}\;\sqrt{1-\frac{P^2}{S'^2}}\Big) }\;, 
\label{20p} 
\end{equation} 
where $S= e_{\frac{1}{2}P+p} + e_{\frac{1}{2}P-p}$  
and $S'= e_{\frac{1}{2}P+p'} + e_{\frac{1}{2}P-p'}$.  
The presence of the factor $(S-S' )^2$ in the numerator indicates that the  
correction, $\Delta$, which has to be removed from the interaction,  
Eqs.~(\ref{20n},~\ref{20o}), to make it independent of  $\vec{P}$, has  
an off-energy shell character. It therefore corresponds to a higher  
order in the interaction. The removing of the $\Delta$ term in the meson 
propagator to fulfill covariance does not therefore spoil the relation of the 
interaction to be used to the field-theory based one in the small coupling 
limit.

The above developments can be extended to  
any form of the desired potential in the c.m. system. While the present  
case has an evident relation to a field-theory motivated interaction,  
it cannot account for its full genuine character. In particular,  
retardation effects have to be mocked up by choosing appropriately the  
form of the interaction (coupling constants, range and non-locality)  
and there is no reason it should be as simple as Eq.~(\ref{20m}). 
 
The Lorentz-invariance property of the interaction, Eq.~(\ref{20n}), is  
lost when the correction, Eq.~(\ref{20p}), is accounted for to recover  
the covariance of the mass spectrum. This is however consistent with  
an instant-form approach while it would not be with a point-form approach  
where a dependence of this interaction on the total momentum, $\vec{P}$,  
should be excluded. We also notice that the interaction, Eq.~(\ref{20n}),  
including the correction, Eq.~(\ref{20p}), is independent of the total  
mass of the system under consideration, $M$, consistently with the  
relativistic-quantum-mechanics character of the present approach.  
This property is absent in the work by Wallace~\cite{Wallace:2001nv}, which  
has some similarity with the present one in a few respects but differs  
in other ones.

A particular interaction of interest is a ``zero-range'' one, corresponding  
to an infinite value for the boson mass, $\mu$. In such a case, the wave  
function, $\Phi(\vec{p}_1,\vec{p}_2)$, is simply given by: 
\begin{equation} 
\Phi(\vec{p}_1,\vec{p}_2)=  
\frac{1}{\sqrt{N}}\;\frac{4\,\pi}{\Big(E_P^2-(e_{p_1}+e_{p_2})^2\Big)}\; 
\frac{\sqrt{2\,(e_{p_1}+e_{p_2})}}{ \sqrt{2\,e_{p_1}} \, \sqrt{2\,e_{p_2}} }\;, 
\label{20qa} 
\end{equation} 
where $N$ is an appropriate normalization. Similarly to the fully  
non-relativistic case with a separable potential in the ``zero-range'' limit,  
this wave function is uniquely   
determined by the mass of the system under consideration. Interestingly, it  
coincides with the one that the examination of the form factor at $Q^2=0$ for a  
Feynman triangle diagram suggests~\cite{Desplanques:2001ze}: 
\begin{equation} 
F_1(0)= 
\frac{16\,\pi^2}{N} \;\int \frac{d\,\vec{p}_1\;d\,\vec{p}_2}{(2\,\pi)^3}  
\; 
\frac{(e_{p_1}+e_{p_2})\;  
\delta(\vec{p}_1+\vec{p}_2-\vec{P})}{2\,e_{p_1}\,e_{p_2}\, 
\Big(E_P^2-(e_{p_1}+e_{p_2})^2\Big)^2}\;. 
\label{20qb} 
\end{equation} 
Assuming the normalization  $F_1(0)=1$, one gets the following expression for  
the normalization factor: 
\begin{equation}   
N=\frac{1}{M^2}\left( \frac{4\,m^2}{M\sqrt{4\,m^2-M^2}}\;{\rm arctg}  
\left(\frac{M}{\sqrt{4\,m^2-M^2}}\right)-1 \right).  
\label{20qc} 
\end{equation} 

It is noticed that the change of variable given by Eq.~(\ref{20e}) is  
accounted for at the lowest $1/m^2$ order by the following boost  
transformation of wave functions in configuration space, which is  
often mentioned in the literature: 
\begin{equation} 
\psi \rightarrow \;e^{i \chi} \psi, \;\; {\rm with} \;\; \chi=  
-\frac{\vec{r} \cdot \vec{P}\;\vec{k} \cdot \vec{P}}{16\,m^2} +h.c. 
 \label{20s} 
\end{equation} 

Finally, one should also notice that Eq.~(\ref{20j}) has a structure very  
similar to a light-front equation for scalar particles~\cite{Carbonell:1998rj}.  
Not surprisingly, this similarity includes the last term in Eq.~(\ref{20o}),  
$\Delta$, defined by Eq.~(\ref{20oo}) 
(in the large $P$ limit, this one reads  $- 
\vec{n} \cdot \vec{k} \;\; \vec{n} \cdot \vec{k}\,' \times  
(e_k-e_{k'})^2/(e_k \,e_{k'})$ if 
one identifies $\vec{P}/P$ with the  light-front orientation, $\vec{n}$).  
However, it does not extend to the mass-dependent term,  which characterizes  
the field-theory foundation of this light-front approach and, evidently, is 
not  part of a quantum mechanics approach.

\section{Form factors} 
In a field-theory based approach, the expressions of the interaction and the  
current that have to be used in the calculation of form factors are generally  
known. In relativistic quantum-mechanics approaches, like the ones considered  
here, further work is necessary to obtain them, requiring a  
procedure to get rid  
of the energy dependence that the field-theory model implies. This can be  
performed in the spirit of the Foldy-Wouthuysen  
transformation~\cite{Amghar:2000pc} or 
of works of refs.~\cite{Fukuda:1954,Okubo:1954}, but  
not much has been done along these lines. The interaction is rather fitted to  
some experimental data like phase shifts or binding energies, keeping the form  
of what field theory suggests. Under these conditions, the determination of the  
current, which contains a single-particle component and a many-body one,  
is more uncertain.  
 
A first possibility for the single-particle current is to take again 
what field  theory suggests, namely the free particle one. While 
doing so, we found that   form factors calculated in the instant-  and  
front-form approaches are identical  in some limit. This result, 
which is not totally unexpected, occurs when the  instant-form 
calculation is made in a configuration where the sum of the  
momenta, $\vec{P}_f+\vec{P}_i~(=2\,\vec{\bar{P}})$, goes to infinity 
while the  momentum transfer  remains perpendicular to this vector. 
We also found that the comparison of  scalar and charge form factors 
evidenced a less pleasant feature in the  strong binding limit. 
At small momentum transfers, their ratio tends to zero  
with $M$, contrary to what more elaborate calculations  
show~\cite{Desplanques:2001ze,Desplanques:2001}.  A detailed analysis 
has revealed that this result points to the omission of  specific 
interaction currents. In the small momentum limit for instance, these  
ones, which mainly affect the charge form factor, combine with a free 
particle  contribution proportional to $2\, m$ to provide a total 
result proportional to  $M$.  Knowing that these interaction currents 
can be cast into the form of a  single-particle but energy-dependent 
current, a different  single-particle current would solve the above 
problem, offering a second  possibility.  
 
\subsection{Definition of form factors in the instant-form approach} 
 
\begin{figure}[htb] 
\begin{center} 
\mbox{\psfig{file=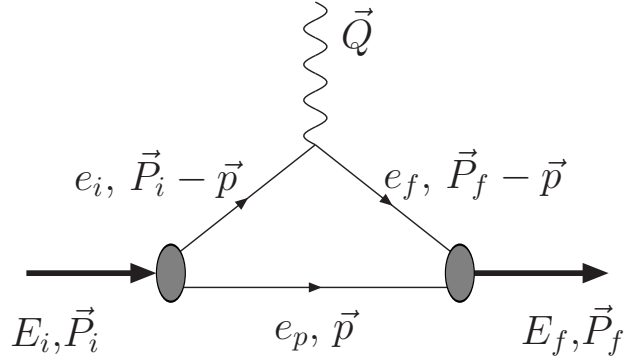,width=20em}}   
\end{center} 
\caption{Diagram showing the interaction with an external probe:  
the kinematical definitions refer to those used for the  
calculation of form factors in an instant-form approach, as described  
in the text. The thick external lines may represent a single-particle  
(triangle Feynman diagram case) or a two-body system (scattering  
on a bound state).}\label{fig1} 
\end{figure}   
   
\subsubsection{Vector form factors} 
For our purpose, we use a current directly suggested by the equation of  
motion, Eq.~(\ref{20a}), where the four-vector dependence on the momenta,  
$P_f^{\mu}+P_i^{\mu}$, is factored out in agreement with current  
conservation for an elastic process (or gauge invariance) and Lorentz  
covariance. In this way, we account implicitly for some interaction  
(two-body) currents. As for the expression of the elastic form factor  
in terms of wave functions, we require that, at $\vec{Q}=0$, it is identical  
to the scalar product ensuring the orthogonality of states with different  
eigenvalues. For an instant-form calculation, the expression of the  
current thus reads: 
\begin{eqnarray} 
\sqrt{2\,E_f\;2\,E_i} \;\langle f|J^{\mu}|i \rangle&=&  
(P_f^{\mu}+P_i^{\mu}) \; F_1(Q^2) 
\nonumber \\ 
&=& (P_f^{\mu}+P_i^{\mu})\int \frac{d\vec{p}}{(2\,\pi)^3} \;  
\Phi(\vec{P}_f-\vec{p},\vec{p}\,) \;  
\Phi(\vec{P}_i-\vec{p},\vec{p}\,), 
\label{30a} 
\end{eqnarray} 
where $\vec{p}$ now refers to the spectator particle (see Fig.~\ref{fig1} 
for a schematical representation of the process and definition of the  
kinematics). Corresponding to this notation, we introduce the  
on-mass shell energies of the interacting constituents in the initial  
and final states, $e_i =\sqrt{m^2+(\vec{P}_i-\vec{p})^2}$  
and $e_f =\sqrt{m^2+(\vec{P}_f-\vec{p})^2}$. At this point, two  
features are to be noticed.  
 
First,  using Eqs.~(\ref{20i}) and (\ref{20l}), one can verify that the form  
factor at zero-momentum transfer as defined by Eq.~(\ref{30a}), $F_1(0)$, is  
independent of the momentum, $\vec{P}$.  This result, which is not a priori  
guaranteed, is in complete agreement with the same independence that we  
require for the mass spectrum  when solving Eq.~(\ref{20a}) in the  
instant-form approach.  
 
Second, for non-zero-momentum transfers, one  
can imagine to introduce an extra factor in the integrand of Eq.~(\ref{30a})  
such as $ (e_f +e_i)/\sqrt{4\,e_f\;e_i}$. This factor, which is  close to 1 
in most kinematical configurations of interest, can be motivated for  some 
part. Examination of the expression of the complete matrix element for a  
``zero-range'' interaction shows however that other factors involving the 
same  order of numerical uncertainty should be considered. These various 
corrections  partly originate from the quantum mechanics character of 
the approach used here,  which prevents one from blindly relying on the 
underlying field-theory model.  Some of them affect the single-particle 
current while other ones, energy  dependent, can be cast into the form 
of two-body currents. All of them, most  often, are required to ensure 
Lorentz covariance or equivalence with other approaches such as an ``exact'' 
calculation of form factors or a different  relativistic quantum mechanics 
approach. In the present work, since we are  mostly interested in the 
gross features of form factors calculated from a  single-particle current 
in different approaches, the many-body contributions  will not be  
considered in detail. Anticipating on later discussions, we  introduce  
the corrections to the single-particle current as follows: 
\begin{eqnarray} 
\sqrt{2\,E_f\;2\,E_i} \;\langle f|J^{\mu}|i \rangle  
= (P_f^{\mu}+P_i^{\mu}) 
\int \frac{d\vec{p}}{(2\,\pi)^3} \;  
\Phi(\vec{P}_f-\vec{p},\vec{p}\,) \;  
\Phi(\vec{P}_i-\vec{p},\vec{p}\,) 
\nonumber \\ \times  
\frac{e_f +e_i}{\sqrt{4\,e_f\;e_i}} \; 
\frac{2\,\sqrt{e_f +e_p}\,\sqrt{e_i +e_p}}{e_f +e_i+2\,e_p}. 
\label{30a2} 
\end{eqnarray} 
The first correction factor has its origin in the elementary coupling  
of a photon to constituents. The second one looks arbitrary at this  
point. Combined with factors present in the $\Phi(\vec{P}-\vec{p},\vec{p})$  
function, it allows one to recover the expression of the diagram of  
Fig.~\ref{fig1} in the limits $E_i=e_i+e_p,\;E_f=e_f+e_p $, which suppresses  
Z-type contributions\footnote{Strictly speaking, Z-type contributions  
and the related concept of backward time-ordered diagrams have no room  
in relativistic quantum mechanics. When referring to them in the  
present paper, we have implicitly in mind the effective contributions  
which account for them in these approaches, such as contact terms for  
instance.}. The overall correction factor is 1 at zero-momentum  
transfer, ensuring that $F_1(0)$, remains independent of the momentum,  
$\vec{P}$.  Relaxing the condition that the contribution of the above  
time-ordered diagram be recovered, one can imagine to introduce extra  
factors in Eq.~(\ref{30a2}), accounting for two-body currents, whose 
interaction character can be transformed away in the case $E_i=E_f$. An  
example of such a factor is suggested by the consideration of the infinite  
boson-mass model (see expression of $F_1(Q^2)$ in Eq.~(\ref{31kc}) below).  
Written as: 
\begin{equation} 
c.f.(E_i=E_f)=\frac{(e_f +e_i+2\,e_p)^2}{(e_f +e_p)\,(e_i +e_p)} \; 
\frac{e_f\;e_i}{(e_f +e_i)^2} \; , 
\label{30a3} 
\end{equation} 
this factor is equal to 1 at zero momentum transfer, ensuring that $F_1(0)$,  
remains independent of the momentum, $\vec{P}$, as above. 
 
\subsubsection{Scalar form factors} 
In calculating form factors relative to a scalar probe, we cannot rely  
on constraints such as current conservation or a minimal coupling  
principle to determine the expression of the operator entering the matrix  
element. It is however reasonable at least to require that the form  
factor at zero-momentum transfer be independent of the total momentum,  
$\vec{P}$, consistently with the similar independence for the spectrum  
of Eq.~(\ref{20a}) in the instant-form approach. Assuming that the form  
factor is the same as the charge one in the non-relativistic limit, one  
immediately gets a possible expression: 
\begin{eqnarray} 
\nonumber 
\sqrt{2\,E_f\;2\,E_i} \;\langle f|S|i \rangle &=& \\ 
4\,m\;F_0(Q^2) &=& 4\,m\; 
\int \frac{d\vec{p}}{(2\,\pi)^3} \; \Phi(\vec{P}_f-\vec{p},\vec{p}\,) \;  
\Phi(\vec{P}_i-\vec{p},\vec{p}\,), 
\label{30b} 
\end{eqnarray} 
which involves the same integral as for the current probe, Eq.~(\ref{30a}).  
 
In comparing the ratio of the scalar and charge form factors obtained  
from Eqs.~(\ref{30a}) and (\ref{30b}) with the one obtained from an  
``exact'' calculation (Wick-Cutkosky model for a zero-mass boson or  
Feynman triangle diagram for an infinite mass), we found it is too small  
at low $Q^2$ and too large at high $Q^2$. The discrepancy is moderate  
and does not go beyond a factor 2. However, when looking at the  
expression of the matrix elements, Eqs.~(\ref{30a}) and (\ref{30b}),  
it is found that the ratio of the scalar and charge form factors is  
not recovered in the non-interacting case at  zero-momentum transfer.  
This ratio, $(e_f+e_f+2\,e_p)/[2\,(e_f+e_i)]$, could be introduced in  
Eq.~(\ref{30b}) but, when doing so, the independence of the scalar  
form factor of the momentum $\vec{P}$ at zero-momentum transfer is lost  
(a factor 1.5 between $P=0$ and $P=\infty$ for the infinite boson-mass  
model). As far as we can see, this can be  partly remedied at the  
expense of introducing an energy dependence in the expression of the  
scalar form factor so that to recover the above ratio in the free-particle  
case. On the other hand, corrections included in Eq.~(\ref{30a2})  
should be accounted for. An improved expression of the scalar matrix  
element is thus given by:  
\begin{eqnarray} 
\label{30c} 
\nonumber 
\lefteqn{ 
\sqrt{2\,E_f\;2\,E_i} \;\langle f|S|i \rangle= 4\,m 
\int  \frac{d\vec{p}}{(2\,\pi)^3}\;\Phi(\vec{P_f}-\vec{p},\vec{p}\,) \;  
\Phi(\vec{P_i}-\vec{p},\vec{p}\,)}  \\ && \qquad \quad 
\times \, 
\frac{\sqrt{e_f +e_p}\,\sqrt{e_i +e_p}}{\sqrt{4\,e_f\;e_i}} \; 
\left( 1+e_p \; \frac{(e_f+e_p)\,(e_i+e_p)-E_f\,E_i} 
{(e_f+e_p)\,(e_i+e_p)\,(e_f+e_i)} \;\right). 
\end{eqnarray} 
This expression is suggested by the analysis of the contribution of the  
Feynman triangle diagram (see expression below) where a tedious calculation  
allows one to check that the form factor at zero momentum transfer does  
not depend on the momentum, $\vec{P}$. In comparison with Eq.~(\ref{30b}),  
it provides some enhancement of the form factor at small $Q^2$ and a  
decrease by roughly a factor 2 at the largest $Q^2$, both in agreement  
with what an ``exact'' calculation suggests. Most of the effect is due to the  
last factor in Eq.~(\ref{30c}). 
 
\subsubsection{Expressions of form factors} 
The expressions of charge and scalar form factors are obtained from 
Eqs.~(\ref{30a2}) and~(\ref{30c}), where wave functions $\Phi(...)$ are 
replaced according to Eq.~(\ref{20l}): 
\begin{eqnarray} 
F_1(Q^2)&=& 
\int \frac{d\vec{p}}{(2\,\pi)^3} \;  
\frac{ \sqrt{ e_{k_{tf}}\,e_{k_{ti}} } }{ e_p }\; 
\phi_0(\vec{k}_{tf}^2) \;  \phi_0(\vec{k}_{ti}^2) \; 
 \frac{(e_f +e_p)\,(e_i +e_p)\,(e_f +e_i)}{2\,e_f\,e_i\,(e_i+e_f+2\,e_p)}, 
\nonumber \\  
F_0(Q^2)&=& \int \frac{d\vec{p}}{(2\,\pi)^3} \;  
\frac{ \sqrt{ e_{k_{tf}}\,e_{k_{ti}} } }{ e_p }\; 
\phi_0(\vec{k}_{tf}^2) \;  \phi_0(\vec{k}_{ti}^2) \; 
\nonumber \\ & & \times  
 \frac{(e_f +e_p)\,(e_i +e_p)}{4\,e_f\,e_i}\; 
 \left( 1+e_p \;\frac{(e_f+e_p)\,(e_i+e_p)-E_f\,E_i} 
{(e_f+e_p)\,(e_i+e_p)\,(e_f+e_i)} \; \right). 
\label{31c} 
\end{eqnarray} 
The arguments in the wave functions, $\vec{k}_{ti}^2$ and  
$\vec{k}_{tf}^2$, read: 
\begin{equation} 
\nonumber 
\vec{k}_{ti,f}^2=\left(\frac{1}{2}\vec{P}_{i,f}-\vec{p}\,\right)^2-   
\left(\frac{\vec{P}_{i,f}\cdot(\frac{1}{2}\vec{P}_{i,f}-\vec{p}\,)} 
{(e_{P_{i,f}-p}+ e_{p} )}\right)^2. 
\label{31d} 
\end{equation} 
The appearance of quantities such as $(e_i +e_p)$ or $(e_f +e_p)$ in 
Eq.~(\ref{31c}) is unusual. It makes sense however if one looks at the  
expression of the wave function $\phi_0(\vec{k}_{ti}^2)$, which involves a term  
$(E_i^2-(e_i +e_p)^2)^{-1}=(E_i-e_i -e_p)^{-1}\;(E_i+e_i +e_p)^{-1}$, see 
Eq.~(\ref{20a}). Up to a factor 2, this extra quantity just cancels the second  
non-singular factor in the limit $E_i \rightarrow e_i +e_p$. As for the quantity  
$(e_f +e_i)/(e_i+e_f+2\,e_p)$ in the expression of $F_1(Q^2)$, it represents the  
ratio of the coupling of the photon to the vector current, a priori expected, to  
the term $E_i+E_f$ which is factored out and is equal to $e_i+e_f+2\,e_p$ in the  
free particle case. Despite the factorization of the quantity  
$(P_f^{\mu}+P_i^{\mu})$ in Eq.~(\ref{30a2}), our expression of the charge form  
factor thus keeps some track of the genuine coupling. Not surprisingly, the  
above ratio is absent in the expression of the form factor relative to a scalar  
probe, $F_0(Q^2)$. 
 
\noindent 
The expressions in the Breit frame are obtained from Eq.~(\ref{31c}) by making  
the replacements $e_i=e_{-\frac{Q}{2}-p}$ and $e_f=e_{\frac{Q}{2}-p}$. 
Specializing here and below to a momentum transfer along the $z$ axis, the  
arguments of the wave functions are then given by: 
\begin{eqnarray} 
\vec{k}_{ti}^2 & = & p^2_x+p^2_y+ (p_z+Q/4)^2 \; \left(1-  
\frac{(Q/2)^2}{(e_{-\frac{Q}{2}-p}+ e_{p} )^2}\right), 
\nonumber \\ 
\vec{k}_{tf}^2 & = & p^2_x+p^2_y+ (p_z-Q/4)^2 \; \left(1-  
\frac{(Q/2)^2}{(e_{\frac{Q}{2}-p}+ e_{p} )^2}\right). 
\label{31f} 
\end{eqnarray} 
We now present the expressions that form factors take in other approaches. 
 
\subsection{Expressions of form factors in other approaches} 
\subsubsection{Point-form calculation} 
For the  form factors in the point form, we start from  currents similar  
to the ones  
used to derive the expressions of the form factors in the instant form, 
Eqs.~(\ref{30a}) and (\ref{30b}). In doing so, we want to avoid 
some major bias in the comparison of results for different 
forms. Beginning with the charge form factor, we notice that each factor  
in the instant-form expression of $F_1(Q^2)$, Eq. (\ref{31c}), 
has an obvious counterpart in a covariant calculation: \vspace{-5mm}
\begin{eqnarray}
 \frac{1}{2\,e} &\rightarrow& \int dp^0 \delta(p^2-m^2) \;
\theta(\lambda_i \cdot p),
\nonumber \\
 2\,\sqrt{e_{k_{tf}}\; e_{k_{ti}} }&\rightarrow& 
\Big((p_f+p)^2 \; (p_i+p)^2\Big)^{1/4} \; 
(=2\,\sqrt{\lambda_f \cdot p \;\; \lambda_i \cdot p}\;),
\nonumber \\
 e_i+e_p  &\rightarrow&  \lambda_i \cdot (p_i+p),
\nonumber \\  
e_f+e_i+2\,e_p  &\rightarrow& (p_f+p_i+2\,p)\cdot(\lambda_i+\lambda_f)\,.
\label{replacements}
\end{eqnarray}
We notice that, in place of the second expression above, a different factor, 
$\;p\cdot(\lambda_i+\lambda_f)$, was used in an earlier version of this work. 
It was introduced in the integrand for the $F_1(Q^2)$ form factor, 
without reference to Eq. (\ref{31c}), so that to ensure  that the current 
be unchanged at $Q=0$. Compared with the factor, 
$(e_f+e_i+2\,e_p)$, it was found to play a 
relatively minor role. The choice adopted in 
Eq. (\ref{replacements}) is better founded however. 
In considering the scalar form factor, $F_0(Q^2)$, we continue to assume that 
the ratio of its integrand to that one for the charge form factor is given by  
$(e_f+e_i+2\,e_p)/[2\,(e_f+e_i)]$.   
This one turns out to be important in order to get the ratio of the scalar and 
charge form factors, $F_0(Q^2)/F_1(Q^2)$, approximately right in the limit of 
large momentum transfers or large binding. The extra correction introduced in 
the instant-form scalar form factor, Eq.~(\ref{30c}),  
in order to make it independent of the total momentum, $\vec{P}_{i,f}$, at zero 
momentum transfer, at least in the infinite-mass boson case, is not needed here 
since the present formalism is  
covariant in any case. The above requirements together with Lorentz invariance 
and the condition of recovering the expression of the free current in the small 
coupling limit are then sufficient to provide a minimal expression for 
the form factors:\vspace{-5mm}

\begin{eqnarray} 
\lefteqn{ 
F_1(Q^2)=  
\frac{1}{(2\pi)^3 } 
\int d^4p \, d^4p_f \,  d^4p_i \, d\eta_f \, d\eta_i \;}  
\nonumber \\ && 
\times \, 
\;\Big( (p_f+p)^2 \, (p_i+p)^2 \Big)^{1/4} \;
\phi_0 \Big((\frac{p_f-p}{2})^2\Big)  \;   
\phi_0 \Big( (\frac{p_i-p}{2})^2 \Big) 
\nonumber \\ && 
\times \, 
\delta(p^2-m^2) \;  \delta(p^2_f-m^2) \; \delta(p^2_i-m^2) \nonumber \, 
\delta^4(p_f+p-\lambda_f \eta_f)  
\;\delta^4(p_i+p-\lambda_i \eta_i) \\ \nonumber && 
\times \, \theta( \lambda_f \cdot p_f) \; \theta(\lambda_f \cdot p)  
\;\theta(\lambda_i \cdot p)\;  
\theta(\lambda_i \cdot p_i)  \, 
\;\;\lambda_f \cdot (p_f+p)\;\;\lambda_i \cdot (p_i+p)\;
\\ && \times \,  \nonumber 
\frac{2\,(p_f+p_i) \cdot (\lambda_f+\lambda_i )}{(p_f+p_i+2\,p) \cdot   
(\lambda_f+\lambda_i )}, 
\nonumber \\  
\lefteqn{ 
F_0(Q^2)=  
\frac{1}{(2\pi)^3 } 
\int d^4p \, d^4p_f \,  d^4p_i \, d\eta_f \, d\eta_i \; } \nonumber\\  
&& \times \,\;\Big( (p_f+p)^2 \, (p_i+p)^2 \Big)^{1/4} \;
\phi_0 \Big((\frac{p_f-p}{2})^2\Big)  \;   
\phi_0 \Big( (\frac{p_i-p}{2})^2 \Big) 
\nonumber \\ && 
\times \, 
\delta(p^2-m^2) \; \delta(p^2_f-m^2) \; \delta(p^2_i-m^2) \nonumber \, 
\delta^4(p_f+p-\lambda_f \eta_f)  
\;\delta^4(p_i+p-\lambda_i \eta_i)  \nonumber \\&& 
\times \, \theta( \lambda_f \cdot p_f) \; \theta(\lambda_f \cdot p)  
\;\theta(\lambda_i \cdot p)\;  
\theta(\lambda_i \cdot p_i) \nonumber \, 
\;\;  \lambda_f \cdot (p_f+p)  \;\;  \lambda_i \cdot (p_i+p)\;.
\label{31pf} \\ 
\end{eqnarray} 
 
Notations, which evidence explicitly the Lorentz invariance, 
have been explained in Ref.~\cite{Desplanques:2001zw}. 
Let us only mention here that the four-vectors $\lambda_{i,f}^{\mu}$ are  
proportional to the total momenta $P_{i,f}^{\mu}$. On the other hand,  
the $\delta(...)$, when integrated over the $\eta_{i,f}$  variables, imply 
relations between momenta similar to Eq.~(\ref{20ba}), but with different  
$\lambda^{\mu}$. The same  $\delta(...)$ functions imply relations such as
$$\eta_i=2\,\lambda_i \cdot p=2\,\lambda_i \cdot 
p_i=\sqrt{(p_i+p)^2}\;,$$\\ 
allowing one to rewrite Eqs. (\ref{31pf}) in various forms. Up to  the 
notations and the last factor introduced in the expression of the 
charge form factor, these 
equations could be obtained from applying methods used 
in Refs. \cite{Allen:2000ge,Wagenbrunn:2000es}. 

In the Breit frame, the above expressions read: 
\begin{eqnarray} 
F_1(Q^2) & =  & \frac{1+v^2}{\sqrt{1-v^2}}\; \int \frac{d \vec{p}}{(2\pi)^3}  
\;\phi_0(\vec{k}_{tf}^2) \;\phi_0(\vec{k}_{ti}^2)\;
\sqrt{1-\Big(\frac{\vec{p}\cdot\vec{v}}{e_p} \Big)^2}\;,  
\nonumber \\ 
F_0(Q^2) & =  &  \frac{1}{\sqrt{1-v^2}}\;\int \frac{d  
\vec{p}}{(2\pi)^3}  
\;\phi_0(\vec{k}_{tf}^2) \;\phi_0(\vec{k}_{ti}^2)\;
\sqrt{1-\Big(\frac{\vec{p}\cdot\vec{v}}{e_p} \Big)^2}\;, 
\label{31g} 
\end{eqnarray} 
where  
\begin{equation} 
\vec{k}_{ti}^2=p^2_x+p^2_y+ \left(\frac{p_z+v \, e_p}{\sqrt{1-v^2}}\right)^2, 
\;\;\;\; 
\vec{k}_{tf}^2=p^2_x+p^2_y+ \left(\frac{p_z-v \, e_p}{\sqrt{1-v^2}}\right)^2, 
\label{31h} 
\end{equation} 
and $v^2=Q^2/(4\,M^2+Q^2)$ for the elastic case considered throughout this  
work. Expressions for the arguments that can be more easily compared to other  
ones in different approaches are as follows: 

\begin{eqnarray} 
\vec{k}_{ti}^2=p_x^2+p_y^2+ \left(\frac{p_z}{\sqrt{1-v^2}}+\frac{Q}{4}\;  
\frac{2\, e_p}{M}\right)^2, \nonumber \\ 
\vec{k}_{tf}^2=p_x^2+p_y^2+ \left(\frac{p_z}{\sqrt{1-v^2}}-\frac{Q}{4}\;  
\frac{2\, e_p}{M}\right)^2. 
\label{31i} 
\end{eqnarray} 
%
 
\subsubsection{Front-form calculation} 
Expressions for the form factors in the front form have been given in 
different places  for the standard configuration $q^+=q^0+q^z=0$.  
We give them here, using the notations of the work by  Karmanov 
and Smirnov~\cite{Karmanov:1992fv} and taking care that the wave  
function entering this calculation  involves an extra factor, 
see Eq.~(\ref{20k}): 
\begin{eqnarray} 
F_1(Q^2) & = & \frac{1}{(2\,\pi)^3}\int  d^2R \int_0^1 \frac{dx}{2x(1-x)} \;  
\tilde{\phi}(\vec{R}^2,x) \; \tilde{\phi}((\vec{R}-x\vec{Q})^2,x) , 
\nonumber \\ 
F_0(Q^2) & = & \frac{1}{(2\,\pi)^3}\int d^2R \int_0^1 \frac{dx}{4x(1-x)^2} \;  
\tilde{\phi}(\vec{R}^2,x) \; \tilde{\phi}((\vec{R}-x\vec{Q})^2,x), 
\label{31j} 
\end{eqnarray} 
where $\tilde{\phi}(\vec{R}^2,x)$ is obtained from  $\tilde{\phi}(\vec{k}^2)$,  
Eq.~(\ref{20k}), by the following replacement of the argument: 

\begin{equation} 
\vec{k}^2= \frac{m^2+\vec{R}^2}{4\,x\,(1-x)}-m^2=\vec{R}^2+ \; 
\frac{m^2+\vec{R}^2}{4\,x\,(1-x)} \;(1-2\,x)^2. 
\label{31k} 
\end{equation} 
The expression of the arguments entering the wave functions in Eq.~(\ref{31j}),  
also given in  Eq.~(\ref{31k}) for a particular case, can be recovered from 
an  equation identical to Eq.~(\ref{20fb}) together with the relation that 
momenta  $\vec{p}_1$, $\vec{p}_2$  and the total momentum $\vec{P}$ of 
the system fulfill  in the front-form approach. This one is given by 
Eq.~(\ref{20ba}), where the  ratio $\vec{\lambda}/\lambda^0$ is 
replaced by a unit vector $\vec{n}$. The  calculation, which is 
somewhat tedious, is performed for a kinematical  configuration where 
$E_i=E_f \; ( \vec{Q} \perp (\vec{P}_i+\vec{P}_f))$ and  
$(\vec{P}_i+\vec{P}_f) \parallel \vec{n}$. The momentum of the interacting  
particle is first replaced in terms of the spectator one, using the 
relation  between momenta defined above. The components of this momentum 
perpendicular and  parallel to $\vec{n}$ are then expressed in terms of 
the $\vec{R}$ and $x$  variables.  A last translation, 
$\vec{R} \rightarrow \vec{R}-x \,\vec{Q}/2$,  allows one to get 
arguments of the wave functions as shown in  Eq.~(\ref{31j}). 
Calculations in a configuration different from the above one are possible 
but rarely mentioned in the literature. They are expected to depend  
significantly on interaction effects and should involve sizeable 
contributions from two-body currents.

\subsubsection{``Exact'' calculations} 
In all present cases, calculations of form factors, that can be  
considered  ``exact'', as far as the implementation of relativity 
is concerned,  are available. They can be obtained with a limited 
amount of work in the zero-mass boson case (Wick-Cutkosky 
model~\cite{Wick:1954eu,Cutkosky:1954ru}) and the  infinite-mass one. 
In the case of a finite-mass boson, much more work is  required. 
Some results have been recently obtained by Sauli and Adam for  
momentum transfers up to $Q^2=6\,m^2$~\cite{Sauli:2001zz,Sauli:2001zy}. 
Obviously, the consideration of scalar particles greatly simplifies 
the task but it also avoids to deal with models that could evidence 
critical values for the coupling constant~\cite{Mandelstam:1955sd}. 
Such a feature could significantly complicate the comparison with 
other approaches. 
 
In the case of the Wick-Cutkosky model, some details about the 
calculation  of form factors and their expressions that we use have 
been given  in Refs.~\cite{Desplanques:2001zw,Desplanques:2000}. 
For the infinite-mass boson case, a few  details can be found 
in Ref.~\cite{Desplanques:2001ze}. The corresponding expressions  
are given by: 
\begin{eqnarray} 
F_1(Q^2) & =  & \frac{1}{N} \;\int_0^1 dx \;(1-x)  
\;\frac{1}{Q\,\sqrt{m^2-M^2\,x(1-x)+Q^2\,x^2/4}}\; 
\nonumber \\  &  & \times \, 
\log 
\left(\frac{\sqrt{m^2-M^2\,x(1-x)+Q^2\,x^2/4}+Q\,x/2}{\sqrt{m^2-M^2\,x(1-x) 
+Q^2\,x^2/4}-Q\,x/2}\right), 
\nonumber \\ 
F_0(Q^2) & = & \frac{1}{N} \;\int_0^1 dx\; \frac{1}{2} \; 
\frac{1}{Q\,\sqrt{m^2-M^2\,x(1-x)+Q^2\,x^2/4}}\; 
\nonumber \\  &  & \times \, 
\log  
\left(\frac{\sqrt{m^2-M^2\,x(1-x)+Q^2\,x^2/4}+Q\,x/2}{\sqrt{m^2-M^2\,x(1-x) 
+Q^2\,x^2/4}-Q\,x/2}\right), 
\label{31kb} 
\end{eqnarray} 
where $N$ is given by Eq.~(\ref{20qc}). Alternative  
expressions can be obtained by integrating  
over the time component of the spectator particle in 
the original Feynman diagram. In the Breit frame, they read: 
\begin{eqnarray} 
\nonumber 
F_1(Q^2)&=&\frac{16\,\pi^2}{N} \int \frac{d\,\vec{p}}{(2\,\pi)^3} \\ && 
\frac{(e_f+e_i+2\,e_p)}{2\,e_p\,(e_f+e_i)\,(E^2-(e_f+e_p)^2)\, 
(E^2-(e_i+e_p)^2)}, 
\nonumber \\ 
\nonumber 
F_0(Q^2)&=&\frac{16\,\pi^2}{N}\int \frac{d\,\vec{p}}{(2\,\pi)^3} \\ && 
\frac{(e_f+e_p)\,(e_i+e_p)\,(e_f+e_i+e_p)-e_p\,E^2}{4\,e_f\,e_i\,e_p\, 
(e_f+e_i)\,(E^2-(e_f+e_p)^2)\,(E^2-(e_i+e_p)^2)}, 
\label{31kc} 
\end{eqnarray} 
with $E=\sqrt{M^2+(\vec{Q}/2)^2},\;  
e_i =\sqrt{m^2+(\vec{p}+\vec{Q}/2)^2},\; e_f  
=\sqrt{m^2+(\vec{p}-\vec{Q}/2)^2}$. 
 
In the limit of a vanishing total mass ($M=0$), analytical expressions of the  
above form factors can be obtained. As they may be useful either to get an  
insight on their asymptotic behavior or, more practically, to remedy the  
difficulty to integrate the most delicate part in Eqs.~(\ref{31kc}) in the  
finite-mass case, they are given here: 
\begin{eqnarray} 
\nonumber 
\lefteqn{ 
F_1(Q^2)_{M=0}=\frac{12\,m^2}{Q^2}\; \Bigg\{ 1-\frac{4\,m^2}{Q^2} }\\ && 
\nonumber \qquad 
 +\left( \log\Big[\sqrt{1+Q^2/(4\,m^2)} + Q/(2\,m)\Big]  
 -\frac{2\,m}{Q} \sqrt{1+Q^2/(4\,m^2)} \right)^2 \Bigg\}, \\  
\lefteqn{ 
F_0(Q^2)_{M=0}=\frac{6\,m^2}{Q^2}\; \left(\log\Big[\sqrt{1+Q^2/(4\,m^2)}+  
Q/(2\,m)\Big] \right)^2.} 
\label{31kd} 
\end{eqnarray} 
 
\subsubsection{Galilean-covariant calculation} 
Expressions of form factors with a Galilean boost are well known. They are  
nevertheless displayed here to better emphasize the differences or similarities  
with expressions involving relativistic boosts. Wave functions used in this  
calculation are obtained from the interaction model, Eq.~(\ref{20m}), the  
normalization factors, $1/\sqrt{e_k}$,  appearing in Eq.~(\ref{20c})  
being replaced by $1/\sqrt{m}$, which is consistent with the Galilean  
character of the boost made here. The general expressions are given by: 
\begin{equation} 
F_1(Q^2)=F_0(Q^2)=\int \frac{d\vec{p}}{(2\,\pi)^3} \;  
\phi_f(\vec{k}_{tf}^2) \;  
\phi_i(\vec{k}_{ti}^2), 
\label{31l} 
\end{equation} 
with 
\begin{equation} 
\vec{k}_{ti}^2=\left(\frac{1}{2}\vec{P}_i-\vec{p}\right)^2, 
\;\;\;\; 
\vec{k}_{tf}^2=\left(\frac{1}{2}\vec{P}_f-\vec{p}\right)^2. 
\label{31m} 
\end{equation} 
Galilean invariance of these form factors holds provided that the boost is 
performed according to the premises of this symmetry, i.e. by discarding  
corrections to the mass due to binding energy and other $1/m$ terms.  
 
In the Breit frame, with $\vec{Q}$ along the $z$ axis, the above arguments  
read: 
\begin{equation} 
\vec{k}_{ti}^2=p^2_x+p^2_y+ \left(p_z+\frac{Q}{4}\right)^2 , 
\;\;\;\; 
\vec{k}_{tf}^2=p^2_x+p^2_y+ \left(p_z-\frac{Q}{4}\right)^2. 
\label{31n} 
\end{equation} 
%
 
\subsection{Relations and comments} 
The expressions of form factors given above for different formalisms have  
been derived independently, keeping however in mind that well  
determined limits should be recovered in some cases.  
 
It is first noticed that expressions for the form factors in the 
instant form allow one  
to recover the front-form ones in the kinematical configuration where the  
average momentum carried by the system, $\vec{\bar{P}}$, goes to infinity,  
which defines a specific axis,  while the momentum transfer,  
$\vec{Q}=\vec{P}_f-\vec{P}_i$, remains orthogonal to the direction so defined.  
The result holds for both charge and scalar form factors and is independent of  
the dynamics. It evidently relies on the various factors introduced in the  
definition of the form factors but also on the precise expression taken by the  
arguments entering the wave function, Eq.~(\ref{31d}), thus providing a check of  
their relevance. 
 
A second relation concerns the infinite-mass boson model (``zero-range''  
interaction). It is found that form factors calculated with the front-form  
expressions are identical to the ones directly obtained from the Feynman  
triangle diagram. This can be shown by inserting in Eq.~(\ref{31j}) the  
appropriate wave function. This one, which can be obtained from Eq.~(\ref{20k})  
together with Eq.~(\ref{20qa}), takes a simple form, $\tilde{\phi}(\vec{k}^2)  
\propto 1/(m^2-M^2/4+\vec{k}^2)$. As there is a reduced dependence on the  
interaction (``zero-range'' one) and on the current 
(contributions of pair-terms  
in other formalisms are suppressed here), the present result represents a  
severe test on the implementation of boost effects. 
 
Our third comment, which complements the previous one, concerns the expression  
taken by the argument of the wave function entering the calculation of form  
factors when boost effects are accounted for. Looking at the transformation of  
the minimal factor $1/(m^2-M^2/4+\vec{k}^2)$, which is part of the c.m. wave  
function in all formalisms, it is found that our instant-form expression 
agrees  
with what the analysis of the Feynman triangle diagram indicates, 
Eq.~(\ref{31kc}). In the Breit  
frame for instance, the above factor for the initial state becomes  
$1/(m^2-M^2/4+\vec{k}_{ti}^2)$ with $\vec{k}_{ti}^2$ given by Eq.~(\ref{31f}),  
while the corresponding factor in the other case reads  
$1/[(e_i+e_p)^2/4-E^2/4]$. Even though the expressions appear quite different,  
one can convince oneself, after some algebra, that they are indeed identical  
(Eq.~(\ref{20fb}) is especially useful). This  
represents another severe test on the implementation of boost effects in the  
calculation of form factors. Until now, we could not find any similar result  
when the boost-transformed $\vec{k}^2$ of the point-form approach, 
Eq.~(\ref{31f}), is used. The appearance of the factor $1/e_p$ in our 
instant-form  
expression of the form factor, Eq.~(\ref{31c}), is also worthwhile to be  
noticed. It is not clear that it has a general character but it appears in the  
expression of the form factor for an infinite-mass boson, Eq.~(\ref{31kc}),  
while it is absent in the point-form expressions. It is essential to  
describe the form factors at large $Q^2$, especially their log dependence.   
 
The fourth and last comment concerns the Lorentz contraction  
effect. As mentioned in the introduction, we do not expect it to be accounted  
for by the simple replacement of the momentum transfer, $Q^2\rightarrow  
Q^2\,(1-v^2)=Q^2/[1+Q^2/(4M^2)]$, leading to a constant form factor at large  
$Q^2$. We nevertheless expect this effect to be around in some limit. 
Examining the expressions for the form factors in the instant form, especially 
the  
boost-transformed $\vec{k}^2$, Eq.~(\ref{31f}), one sees that the quantity,  
$(p_z \pm Q/4)^2$, involves an extra factor which is absent for a Galilean  
boost and is close to the factor $1-v^2$. In this limit and taking into 
account  
that an extra factor $\sqrt{1-v^2}$ appears in the integrand of the form 
factor,  
one can reproduce the above recipe. However, the identity between the factor   
$1-v^2$ and the extra factor only holds in the low $Q^2$ and small binding  
limits. At high $Q^2$, the former one tends to zero like $1/Q^2$ while the  
latter one either goes to zero like $1/Q$ (small momentum for the spectator  
particle) or to a constant (longitudinal momentum of the order $p_z \pm Q/4$). 
This is sufficient to make the form factor decrease when $Q^2$ increases.  
The erroneous character of the above recipe can be seen as follows. The 
constant form factor to which it leads at high $Q^2$ implies that the system 
in the Breit-frame has no depth along the momentum transfer. In other words, 
the two constituents arrive and go back together, which is not what occurs 
physically. The photon momentum is transferred to one constituent while the 
other one is not affected. This difference in the role of the two particles is 
accounted for by the departure from the factor $v^2$ evidenced by the last 
term in the two expressions of Eq.~(\ref{31f}), which, moreover, differ from 
each other. Such a departure was found in a numerical study 
by Gl\"ockle and Nogami at the lowest order in the  
interaction~\cite{Glockle:1987hb}. 
However, contrary to their conjecture that this should vanish at higher order, 
we believe that these differences are there and are essential to get the 
appropriate asymptotic behavior of form factors.  
 
\section{Presentation of the results} 
We present here results for elastic form factors calculated with the different  
expressions for the boost transformation given in the previous section. They  
concern the ground state of a two-body system  composed of scalar particles  
interacting by exchanging a scalar boson. Most results have a covariant  
character, Lorentzian or Galilean. Only those for the instant-form approach do  
not have such a property. Their frame dependence will be discussed in  
the second part of this section.  
 
\noindent 
\subsection{Covariant and Breit-frame calculations} 
Numerical results are successively given in three tables:  Table~\ref{t10} for  
a zero-mass boson, Table~\ref{t20} for a finite mass boson ($\mu=0.15\,m$) and  
Table~\ref{t30} for an infinite one. In every case, two masses of the total  
system are considered corresponding to a moderately and a strongly bound  
system, roughly $M\simeq1.6\,m$ and $M=0.1\,m$, respectively (the precise  
values are given in the table).

\begin{table}[tb!] 
\caption{Elastic scalar- and vector-  form factors, $F_1(Q^2)$ and  $F_0(Q^2)$:  
Results for a mass-less boson and two different total  masses of the system,   
$M=1.568\,m$ and $0.1\,m$ ($\kappa^2=0.385\,m^2$ and $\kappa^2=0.9975\,m^2$).  
The table successively contains results  for the instant-form boost  
(I.F.), for the front-form boost (F.F.), and the point-form boost (P.F.), all of  
them with wave function obtained from Eqs.~(\protect\ref{20c})  
and~(\protect\ref{20m}), for the  
Wick-Cutkosky model (B.S.), and for a Coulombian wave function with a Galilean  
boost (Gal.).  
The corresponding coupling constants are given in Table~\protect\ref{t40}. 
Asymptotic behaviors for $F_1(Q^2)$ are $ Q^{-4}\;\log Q$,  
$Q^{-4}\;\log Q$, $Q^{-8}$, $Q^{-4}\;\log Q$, and $Q^{-4}$  
for I.F., F.F., P.F., B.S. and Gal., respectively. Except for a few  
cases, they are not reached in the range of momentum transfers shown in the  
table. 
\label{t10} } 
\medskip 
\begin{center} 
\begin{tabular}{lcccccc} 
\hline  \rule[0pt]{0pt}{3ex} 
  $Q^2/m^2$             &   0.01   &   0.1  &  1.0   &  10.0 & 100.0 
  \\ [1.ex] \hline 
 $M=1.568\,m$                       &          &        &        &       &   \\ 
 $F_1\;\;\;\;$I.F.  & 0.996  & 0.956  & 0.668 & 0.111 & 0.342-02  \\ [0.ex] 
 $F_0\;\;\;\;$I.F.  & 1.237  & 1.182  & 0.787 & 0.106 & 0.253-02  \\ [0.ex] 
 $F_1\;\;\;\;$F.F.  & 0.995  & 0.954 & 0.657  & 0.102  & 0.302-02 \\ [0.ex] 
 $F_0\;\;\;\;$F.F.  & 1.062  & 1.014 & 0.675  & 0.091 & 0.222-02 \\ [0.ex] 
 $F_1\;\;\;\;$P.F.  & 0.993  & 0.928  & 0.507  & 0.195-01 & 0.167-04 \\ [0.ex] 
 $F_0\;\;\;\;$P.F.  & 0.992  & 0.919  & 0.464  & 0.130-01 & 0.087-04 \\ [1.ex] 
 $F_1\;\;\;\;$B.S.   & 0.996   & 0.962  & 0.705  &  0.139  & 0.50-02  \\ [0.ex]  
 $F_0\;\;\;\;$B.S.   & 1.123   & 1.080  & 0.767  &  0.132  & 0.39-02 \\ [1.ex]  
 $F_1=F_0\;{\rm Gal.}$   & 0.997   & 0.968  & 0.740  &  0.145  & 0.337-02 \\  
[1.ex]  \hline 
 $M=0.1\,m$                       &          &        &        &       &   \\ 
 $F_1\;\;\;\;$I.F.  & 0.998 & 0.979 & 0.818 & 0.270 & 0.156-01  \\ [0.ex] 
 $F_0\;\;\;\;$I.F.  & 1.496 & 1.456 & 1.138 & 0.279 & 0.116-01 \\ [0.ex] 
 $F_1\;\;\;\;$F.F.    & 0.998  & 0.977  & 0.804 &  0.251 & 0.142-01 \\ [0.ex]  
 $F_0\;\;\;\;$F.F.    & 1.120  & 1.092  & 0.866 &  0.227 & 0.102-01 \\ [0.ex]  
 $F_1\;\;\;\;$P.F. & 0.431 & 0.806-02 & 0.393-05 & 0.682-09 & 0.95-13 \\ [0.ex] 
 $F_0\;\;\;\;$P.F. & 0.360 & 0.470-02 & 0.200-05 & 0.342-09 & 0.47-13 \\ [1.ex] 
 $F_1\;\;\;\;$B.S.   & 0.998  & 0.983   & 0.848  &  0.338  & 0.283-01 \\ [0.ex]  
 $F_0\;\;\;\;$B.S.   & 1.247  & 1.222   & 1.016  &  0.338  & 0.217-01 \\ [1.ex]  
 $F_1=F_0\;{\rm Gal.}$   & 0.999   & 0.988  & 0.886  &  0.378  & 0.189-01 \\  
[1.ex]  
\hline \\ 
\end{tabular} 
\end{center} 
\end{table}

\begin{table}[tb!] 
\caption{Elastic scalar- and vector-  form factors, $F_1(Q^2)$ and  $F_0(Q^2)$:  
Same as for Table~\ref{t10} but for a  boson mass, $\mu=0.15\,m$. The two total  
masses are  $M=1.6\,m$ and $0.1\,m$. The corresponding coupling constants are  
$g^2/4\pi=1.915$ and $3.392$ for the interaction model given by Eqs.  
(\ref{20c})  
and (\ref{20m}), and  $g^2/4\pi=1.487$ and $2.289$ for model the simple Yukawa  
potential. Bethe-Salpeter results are from 
refs.~\protect\cite{Sauli:2001zz,Sauli:2001zy} (see comments in the  
text).  
\label{t20} } 
\medskip 
\begin{center} 
\begin{tabular}{lcccccc} 
\hline  \rule[0pt]{0pt}{3ex} 
  $Q^2/m^2$             &   0.01   &   0.1  &  1.0   &  10.0 & 100.0 
  \\ [1.ex] \hline 
 $M=1.6\,m$                       &          &        &        &       &   \\ 
 $F_1\;\;\;\;$I.F.  & 0.997  & 0.968  & 0.741  & 0.171 & 0.68-02  \\ [0.ex] 
 $F_0\;\;\;\;$I.F.  & 1.241  & 1.198  & 0.877  & 0.164 & 0.50-02  \\ [0.ex] 
 $F_1\;\;\;\;$F.F.    & 0.997  & 0.966  & 0.729 & 0.158 & 0.61-02 \\ [0.ex]  
 $F_0\;\;\;\;$F.F.    & 1.084  & 1.046  & 0.762 & 0.141 & 0.44-02 \\ [0.ex]  
 $F_1\;\;\;\;$P.F. & 0.995  & 0.948  & 0.608  & 0.381-01 & 0.41-04 \\ [0.ex] 
 $F_0\;\;\;\;$P.F. & 0.994  & 0.939  & 0.5580  & 0.255-01 & 0.22-04 \\ [1.ex] 
 $F_1\;\;\;\;$B.S.   & (0.996)& (0.961) & (0.697) &(0.119)&(0.27-02)  \\ [1.ex]  
 $F_1=F_0\;{\rm Gal.}$   & 0.998 & 0.976 & 0.796 &  0.209  & 0.63-02  \\  
[1.ex] \hline 
 $M=0.1\,m$                       &          &        &        &       &   \\ 
 $F_1\;\;\;\;$I.F.    & 0.999  & 0.984 & 0.859  & 0.349 & 0.263-01  \\ [0.ex] 
 $F_0\;\;\;\;$I.F.    & 1.496  & 1.464 & 1.201  & 0.366 & 0.195-01 \\ [0.ex] 
 $F_1\;\;\;\;$F.F.    & 0.998  & 0.982  & 0.844 &  0.324 & 0.241-01 \\ [0.ex]  
 $F_0\;\;\;\;$F.F.    & 1.150  & 1.126  & 0.931 &  0.298 & 0.173-01 \\ [0.ex]  
 $F_1\;\;\;\;$P.F. & 0.509 & 0.137-01 & 0.76-05 & 0.137-08 & 0.19-12 \\[0.ex] 
 $F_0\;\;\;\;$P.F. & 0.425 & 0.080-01 & 0.39-05 & 0.069-08 & 0.10-12 \\ [1.ex] 
 $F_1=F_0\;{\rm Gal.}$   & 0.999 & 0.990 & 0.908  & 0.451   & 0.293-01    \\  
[1.ex]  
\hline \\ 
\end{tabular} 
\end{center} 
\end{table}  
 
\begin{table}[tb!] 
\caption{Elastic scalar- and vector-  form factors, $F_1(Q^2)$ and  $F_0(Q^2)$:  
Same as for Table~\ref{t10} but for an infinite-mass boson. The two total masses 
are  $M=1.6\,m$ and $0.1\,m$ ($\kappa^2=0.36\,m^2$ and  
$\kappa^2=0.9975\,m^2$). The wave function used in the I.F., F.F. and P.F.  
cases is issued from Eq.~(\protect\ref{20c}). 
Asymptotic behaviors for $F_1(Q^2)$ are $Q^{-2}\;(\log Q)^2$,  
$Q^{-2}\;(\log Q)^2$, $Q^{-4}$, $Q^{-2}\;(\log Q)^2$, and $Q^{-1}$  
for I.F., F.F., P.F., B.S. and Gal., respectively. Except for a few  
cases, they are not reached for the range of momentum transfers shown in the  
table. Part of the results were presented in~\protect\cite{Desplanques:2001ze}. 
\label{t30} } 
\medskip 
\begin{center} 
\begin{tabular}{lcccccc} 
\hline  \rule[0pt]{0pt}{3ex} 
  $Q^2/m^2$             &   0.01   &   0.1  &  1.0   &  10.0 & 100.0 
  \\ [1.ex] \hline 
 $M=1.6\,m$                       &          &        &        &       &   \\ 
 $F_1\;\;\;\;$I.F.   & 0.999   & 0.990  & 0.917  &  0.594  & 0.208  \\ [0.ex]  
 $F_0\;\;\;\;$I.F.   & 1.325   & 1.309  & 1.176  &  0.658  & 0.187  \\ [0.ex]  
 $F_1\;\;\;\;$F.F.   & 0.999   & 0.989  & 0.908  &  0.566  & 0.191  \\ [0.ex]  
 $F_0\;\;\;\;$F.F.   & 1.325   & 1.309  & 1.176  &  0.659  & 0.187  \\ [0.ex]  
 $F_1\;\;\;\;$P.F. & 0.999   & 0.986  & 0.871  &  0.353 & 0.207-01 \\ [0.ex]    
 $F_0\;\;\;\;$P.F. & 0.998   & 0.976  & 0.800  &  0.236 & 0.108-01 \\ [1.ex]     
 $F_1\;\;\;\;$B.S.   & 0.999   & 0.989  & 0.908  &  0.566  & 0.191  \\ [0.ex]  
 $F_0\;\;\;\;$B.S.   & 1.325   & 1.309  & 1.176  &  0.659  & 0.187  \\ [1.ex]  
 $F_1=F_0\;$Gal.   & 0.999   & 0.994  & 0.947  &  0.699  & 0.320  \\ [1.ex] 
 \hline 
 $M=0.1\,m$                       &          &        &        &       &   \\ 
 $F_1\;\;\;\;$I.F.  & 0.999   & 0.996  & 0.963  &  0.759  & 0.343  \\ [0.ex]  
 $F_0\;\;\;\;$I.F.  & 1.498   & 1.487  & 1.389  &  0.920  & 0.320  \\ [0.ex]  
 $F_1\;\;\;\;$F.F.  & 0.999   & 0.995  & 0.954  &  0.723  & 0.315  \\ [0.ex]  
 $F_0\;\;\;\;$F.F.  & 1.498   & 1.487  & 1.389  &  0.920  & 0.320  \\ [0.ex]  
 $F_1\;\;\;\;$P.F. & 0.839  & 0.222   & 0.82-02 & 0.141-03 & 0.20-05 \\[0.ex]    
 $F_0\;\;\;\;$P.F. & 0.699  & 0.130   & 0.42-02 & 0.071-03 & 0.10-05 \\[1.ex]     
 $F_1\;\;\;\;$B.S.  & 0.999   & 0.995  & 0.954  &  0.723  & 0.315  \\ [0.ex]  
 $F_0\;\;\;\;$B.S.  & 1.498   & 1.487  & 1.389  &  0.920  & 0.320  \\ [1.ex]  
 $F_1=F_0\;$Gal.  & 0.999   & 0.998  & 0.980  &  0.846  & 0.475  \\ [1.ex] 
\hline \\ 
\end{tabular} 
\end{center} 
\end{table} 
 
\begin{table}[htb] 
\caption{Mass spectrum for the interaction used together with a zero-boson  
mass:  
Results are presented for two total masses of the system under  
consideration (in units of the constituent mass $m$). They are given for the 
simplest interaction, Eq.~(\protect\ref{20m}), for an improved interaction,   
Eq.~(\ref{40m}), for the Coulombian model used together with a Galilean boost,  
and for the ``exact'' calculation (Wick-Cutkosky model). The various columns 
contain successively the coupling constant, the total mass for the ground state 
(input) and those for the first-radial excitation and $l=1$ states. 
\label{t40} } 
\medskip 
\begin{center} 
\begin{tabular}{lccccc} 
\hline  \rule[0pt]{0pt}{3ex} 
  interaction &$\;\;g^2/4\pi\;\;\;\;$ & $\;l= 0\;$ & $l= 0^*$ & $l=1$ \\ [1.ex]  
  \hline 
 $M=1.568\,m$  &          &              &        &         \\ 
 simplest      & 1.497    &   1.568      & 1.888  & 1.875   \\ [0.ex]   
 improved      & 1.314    &   1.568      & 1.897  & 1.891   \\ [0.ex] 
 Coulombian    & 1.241    &   1.568      & 1.901  & 1.901   \\ [0.ex] 
 ``exact''     &  3       &   1.568    & 1.896  & 1.896  \\ [1.ex] \hline 
 $M=0.1\,m$    &             &       &         &         \\ 
 simplest      & 2.750       & 0.10  & 1.668   & 1.615   \\ [0.ex] 
 improved      & 2.267       & 0.10  & 1.707   & 1.671   \\ [0.ex] 
 Coulombian    & 1.997       & 0.10  & 1.733   & 1.733   \\ [0.ex] 
 ``exact''     & 1.996$\,\pi$  & 0.10  & 1.716   & 1.716   \\ [1.ex]  
\hline \\ 
\end{tabular} 
\end{center} 
\end{table} 
 
In each table, results employing the same wave function but different  
boost transformations, successively instant-form (I.F.), front-form (F.F.)  
and point-form (P.F.), are presented. They are followed by what can be  
considered as an ``exact'' calculation, and  a  
non-relativistic one, characterized by a Galilean boost.  
 
As mentioned in Sect. 2, instant-form, front-form and point-form calculations 
can be performed from a unique wave function, but with appropriate boost 
transformations. Being interested in the ground state of a two-body system, it 
is natural to require that the binding energy of this state be reproduced. 
This determines the unique parameter which enters an interaction like 
Eq.~(\ref{20m}), namely the coupling constant which is given in Table~\ref{t40} 
together with other data. This interaction model is used for 
results presented in Tables~\ref{t10} and~\ref{t20}. As will be seen below, 
this provides in some cases (I.F. and F.F.) results that are not too far from 
the ``exact'' ones. A possibility to improve them in relation with a better 
account of the spectrum of our ``exact'' model will be discussed at 
the end of the section.   
 
Concerning ``exact'' calculations (B.S.), there are some results in the  
zero-mass case   (Wick-Cutkosky model). These ones together with the  
relevant expressions for the form factors have been obtained  
elsewhere~\cite{Desplanques:2001zw,Desplanques:2000}.  
They correspond to the couplings $\alpha=3$ and  
$\alpha=1.996\,\pi$, which lead to the total mass  $M=1.568\,m$ and $M=0.10\,m$.  
In this case, we could have considered the very extreme  case of a  
zero mass, $\alpha=2\,\pi$.  
However, while most results are not sensitive to  the precise  value  
(0.5\% at the largest $Q^2$ values considered here), those for the  
point-form approach are, requiring to take a non-zero  finite value.   
These mass values are the ones used in the fitting of the 
interaction strength  in the other models. There are also ``exact'' 
calculations in the infinite-mass boson case. Results are uniquely  
determined by the knowledge of the mass of  the system, $M$. Finally,  
tentative results, given in parentheses in Table~\ref{t20}, have been recently  
obtained by Sauli and Adam in the 
finite-mass boson cases ($F_1$, $M=1.6\,m$)~\cite{Sauli:2001zz,Sauli:2001zy}.  
Comparing these results with the instant- or front-form  
ones shows that they are relatively lower than what a similar comparison  
for our zero or infinite boson-mass calculation suggests. Examination  
of the method employed by these authors indicates that the interaction is  
implicitly cut off at high momentum, explaining the above observation.

Concerning the non-relativistic case, the wave function is obtained from an  
interaction model ignoring relativistic normalization factors $\sqrt{m/e_p}$.  
Such an interaction with a zero-mass boson (Coulombian potential) reproduces  
well the spectrum of the Wick-Cutkosky model~\cite{Amghar:2000pc}. The bulk of 
the  
effective coupling is understood and only minor adjustments are needed to fit  
the exact binding energy. The corresponding form factors scale like $Q^{-4}$ at  
high $Q^2$ and offer the advantage of taking an analytical expression: 
\begin{equation} 
F_1(Q^2)=F_0(Q^2)=\frac{1}{(1+Q^2/16\kappa^2)^2}, 
\label{40a} 
\end{equation} 
where $\kappa^2=m^2-M^2/4$ (the same definition is referred to in  
Tables ~\ref{t10}-\ref{t30}).  
 
Finally, for each entry, we give results for the charge and scalar form  
factors, $F_1(Q^2)$ and $F_0(Q^2)$, which generally differ from each other.  
This can provide a deeper insight on the role of various ingredients 
introduced in the calculation. 
 
In comparing the various results, we first consider the instant-form (I.F.), 
the front-form (F.F.) and the ``exact'' (B.S.) results. The 
reason is that they evidence most often the same trends. Beyond the gross 
features, the interest  is in the remaining discrepancies at the level of a 
factor 2.  
 
For the general features, we notice that they have the same asymptotic 
behavior, $Q^{-4}$ for the finite-range interaction and $Q^{-2}$ for the 
infinite range one. This is not always transparent from the results presented 
in the tables and, to verify it, one has to go to higher momentum transfers.  
As is seen in the particular case of the Galilean boost applied to a  
Coulombian-type problem, for which the analytical expression of the  
form factor is known, Eq.~(\ref{40a}), the convergence to the asymptotic 
behavior is rather slow, confirming results obtained for instance for the 
three-quark system in ref.~\cite{Desplanques:2000ev}.  
A refined analysis shows that there 
are corrections to this asymptotic behavior with a log dependence that  
enhances the form factor at high $Q^2$. Again, it is not easy to see it  
from examining the tables. The effect is actually responsible, in the  
finite-range interaction case, for the increase of the relativistic  
calculations (I.F., F.F. and B.S.) relative to the Galilean ones  
when going from $Q^2=10\,m^2$ to $Q^2=100\,m^2$. These results show  
the adequacy of the various factors and boost transformations  
involved in the different calculations.  
 
The second general feature concerns the ratio of the form factors, $F_0$  
and $F_1$. Compared to 1, it is larger at low $Q^2$, and smaller at  
the highest values. This results from the choice of the current,  
especially the introduction of a factor in Eq.~(\ref{30c}) to account  
for the ratio of these form factors for the non-interacting case. Again,  
this is a support for the way form factors are calculated (wave functions  
and currents). At the largest values of $Q^2$, a ratio $1/2$ is expected.  
The dependence on log terms, which varies quite slowly, explains why this  
asymptotic result is hardly seen in the tables. Their effect can  
nevertheless be checked for instance on Eq.~(\ref{31kd}). 
 
Besides these general common features, a detailed examination shows  
a few departures. The ratio, $F_0/F_1$, at low $Q^2$ for the instant-form  
calculation is larger than what is expected from the ``exact'' one  
while the front-form calculation shows the opposite trend (finite-range  
interaction case). This indicates that the factor introduced in  
Eq.~(\ref{30c}), taken from the infinite-range interaction case, 
misses something. Notice that in absence of this factor, the ratio  
in the instant-form case  would also disagree but would be smaller  
than the ``exact'' result at low $Q^2$. The truth is probably in  
between. On the other hand, the form factors in the instant and front forms  
tend to be smaller than the ``exact'' ones at high $Q^2$. This points  
to the choice of the effective interaction, which has not been  
optimized and will be discussed below. 
 
It is worthwhile to notice that the front-form calculations generally  
compare well with the  exact results. As is well known, the front-form  
approach ignores Z-type contributions that are implicitly included in  
the formalism. In absence of such contributions, the above agreement  
is not a surprise. This result is useful however, as there are not always  
``exact'' results for the exchange of a finite-mass boson.  
The front-form calculations can then provide benchmark values.

From examining results presented in Tables~\ref{t10}-\ref{t30}, it is evident  
that the point-form calculations of form factors differ strikingly from all the  
other ones, especially in cases where a relativistic treatment is a priori  
required. Their fall-off is much faster and, moreover, it evidences a variation 
with the mass of the system opposite to all the other calculations.  
At small $Q^2$, the slope of the form factors in the point form suggests a  
charge or (Lorentz-) scalar radius significantly larger than in the other cases.  
As an analytical calculation in a particular model (Coulombian one)  
shows~\cite{Desplanques:2001zw}, this  
radius contains a term proportional to $1/M$, hence a radius tending to  
infinity when the mass $M$ goes to zero. One can also guess that the 
finite-mass-boson results  
presented here evidence a power-law behavior   
like $Q^{-8}$ at high $Q^2$, for the charge and  scalar form factors.  
Such behaviors disagree with what is expected from the Born amplitude, which  
provides a $Q^{-4}$ power law. This one is approximately verified in the other  
cases. The difference is entirely due to the way the boost transformation is 
implemented in the wave function. At the places where the ``exact'', 
instant-form and front-form calculations predict some $Q^2$ power law, 
the point-form calculations provide a $ Q^4$ one. This property 
is probably related to the observation made in Ref. \cite{Allen:2000ge} 
that the momentum transfer at the interaction vertex 
with the external probe in the point-form approach, 
$(\vec{p}_i-\vec{p}_f)^2$, does not vary 
like $Q^2$ as generally expected but rather like $Q^2 \;(1+Q^2/(4\,M^2))$. 
While this affects the size of the form factors, it is noticed 
that the ratio of the charge and scalar form factors, $F_1(Q^2)/F_0(Q^2)$, 
turns out to be roughly correct, as a consequence of the extra factor 
introduced in the expression of $F_1(Q^2)$ in Eq. (\ref{31pf}).  
Despite some sizable differences with results presented  
elsewhere~\cite{Desplanques:2001zw,Desplanques:2001ze},  
due to the inputs (currents and wave functions), the most 
important drawbacks, in the large $Q^2$ limit or 
in the $M \rightarrow 0$ limit, persist. 
 
Galilean boost results, mainly given here as a simple benchmark, have already 
been mentioned. Contrary to what one could naively expect, they   
do rather well in the finite-range-interaction cases. This agreement   
is partly due to  the fact that the current is conserved   
and that the Born amplitude constraint is fulfilled. The slight departure to 
the ``exact'' result in the zero-mass-boson case that appears at the highest  
value of $Q^2$ is essentially due to the log terms discussed above. In the 
infinite-mass-boson case, the relatively good agreement of the  Galilean  
boost results with the other ones should be corrected by the observation  
that the asymptotic behavior differs, $Q^{-1}$ against $Q^{-2}\;(\log\,Q)^2$.  
This difference is due to a slower convergence of the integrals entering the 
calculation of the form factors. In comparison,  
the failure of the point-form results, which a priori should improve  
upon these non-relativistic calculations, appears as quite 
striking.

\subsection{Frame dependence of the form factors in the instant form} 
\begin{figure}[htb] 
\begin{center} 
\mbox{\psfig{file=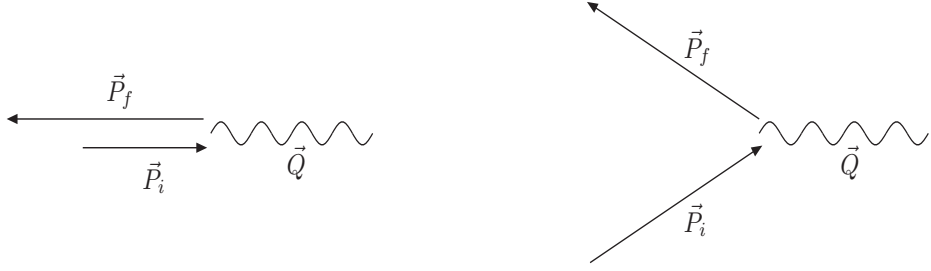,width=0.9\textwidth,height=20ex}}   
\end{center} 
\caption{The left and right parts of the figure represent kinematical  
configurations obtained when giving some extra momentum to the system, 
respectively  parallel  and transverse to the momentum that the initial 
and final states have in the  Breit-frame.}\label{fig2} 
\end{figure}

The instant-form calculation of form factors is not covariant, though it  
used wave functions obtained from an equation that has this property for the  
mass spectrum. To recover a covariant form factor, the contribution of two-body  
currents is generally needed. This can be checked on the example of the triangle  
diagram already mentioned. In absence of such currents, whose derivation is not  
straightforward, the form factors in the instant form can depend on the frame in 
which 
they are calculated. Two particular frames are of interest. One assumes  
a boost along the momentum transfer, which includes in particular the Lab-frame  
case. The other one assumes a boost perpendicular to the momentum transfer (see  
Fig.~\ref{fig2} for a graphical representation). In the limit where an infinite  
momentum is given, it is generally expected that one should recover the 
front-form  results. The two situations are successively considered now.   
 
\begin{figure}[tb] 
\begin{center} 
\mbox{\psfig{file=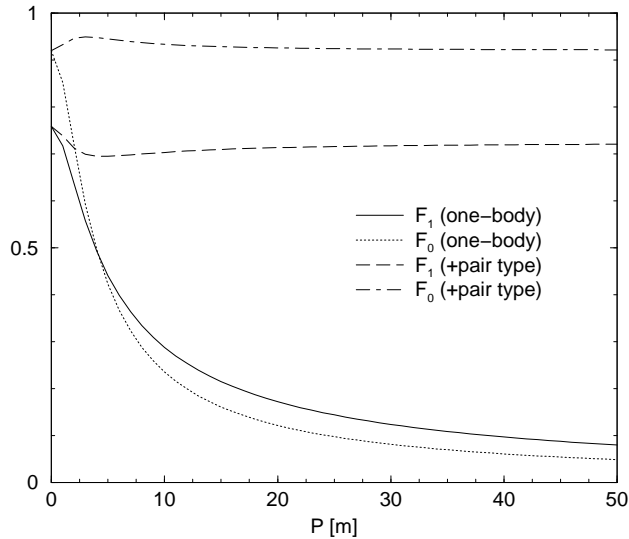,width=20em}}   
\end{center} 
\caption{Dependence of form factors on the average momentum  
$\vec{\bar{P}}$  in the parallel configuration  for $Q^2=10\,m^2$:  
the solid and dotted curves represent the single-particle contribution  
to  $F_1$ and $F_0$,  Eq.~(\protect\ref{31c}), while the dashed and dot-dashed  
curves include the dominant interaction effect, Eq.~(\protect\ref{40b}).} 
\label{fig3} 
\end{figure}

\noindent 
$\bullet$ For the parallel configuration, where, together with the momentum  
transfer, there is a non-zero energy transfer ($E_i \neq E_f$), it has been  
found that the form factors tend to decrease when going away from the  
Breit-frame limit with a possible saturation when the average momentum  
$\vec{\bar{P}}$ goes to $\infty$ (see Fig.~\ref{fig3} for an example in  
the infinite  
boson-mass case). The power-law behavior of the form factors does not  
seem to be affected. As to the reduction, it appears to  increase while  
the total mass $M$ tends to zero. Although presented differently, the  
strong sensitivity of the form factors to the average momentum carried by  
the system has a close relation with the sensitivity to the orientation  
of the quantization surface found in ref.~\cite{Szczepaniak:1995mi}.  
The difference  
with the Breit-frame results is obviously due to two-body currents,  
which actually can take the form of a single-particle but mass-dependent  
current. This can be seen on  the form factors relative to the Feynman  
triangle diagram where analytic expressions are available. The main  
correction to Eqs.~(\ref{31c}) arises from a pair-type diagram and is  
given by:  
\begin{eqnarray} 
\nonumber 
\lefteqn{ 
\Delta F_1(Q^2)= 
\int \frac{d\vec{p}}{(2\,\pi)^3} \;  
\frac{  \sqrt{ e_{k_{tf}}\;e_{k_{ti}} }  }{e_p}\; 
\phi_0(\vec{k}_{tf}^2) \; \phi_0(\vec{k}_{ti}^2) } \\ && \times  
\frac{(E_i-E_f)\,(e_i-e_f)\,(E_i+e_i+e_p)\,(E_f+e_f+e_p)}{8\,e_i\,e_f\,(E_i+E_f) 
(e_i+e_f)} 
\nonumber \\ \nonumber 
\lefteqn{ \times  
\frac{(E_f-e_f-e_p)(E_f-E_i-e_i-e_f)+(E_i-e_i-e_p)(E_i-E_f-e_i-e_f)}{(e_i+e_f)^2 
-(E_i-E_f)^2}, } \\  
\nonumber 
\lefteqn{ 
\Delta F_0(Q^2)= \int \frac{d\vec{p}}{(2\,\pi)^3} \;  
\frac{ \sqrt{ e_{k_{tf}}\;e_{k_{ti}} } }{e_p}\;   
\phi_0(\vec{k}_{tf}^2) \;  \phi_0(\vec{k}_{ti}^2) } \\ && \times  
\frac{(E_i-E_f)\,(E_i+e_i+e_p)\,(E_f+e_f+e_p)}{16\,e_i\,e_f\,(e_i+e_f)} 
\nonumber \\  
\nonumber 
\lefteqn{\times  
\frac{(E_f-e_f-e_p)(E_f-E_i-e_i-e_f)-(E_i-e_i-e_p)(E_i-E_f-e_i-e_f)}{(e_i+e_f)^2 
-(E_i-E_f)^2}. } \\ 
\label{40b} 
\end{eqnarray} 
The interaction dependence of these contributions is made clear if one  
notices that the factors $E_i-e_i-e_p$ or $E_f-e_f-e_p$ at the numerator  
of the last factor, when multiplying the wave function, can be transformed  
away using the mass equation, Eq.~(\ref{20a}). Due to extra E terms,  
the actual dependence on the interaction is much more complicated than  
a linear one. In particular, the term $(E_i-E_f)^2$ at the denominator  
of this last factor is essential. At large $\vec{\bar{P}}$, it tends  
to cancel the other term in the denominator, $(e_i+e_f)^2$, enhancing  
the corresponding overall contributions. This will be lost in a limited  
expansion in terms of the energies. When these contributions are added  
to the single-particle current contribution,  
Eqs.~(\ref{30a},~\ref{30c}), some stability of the results on the average  
momentum $\vec{\bar{P}}$ is recovered.  
This can be seen on Fig.~\ref{fig3} where  
form factors,   $F_1$ and $F_0$, are shown for $Q^2=10\,m^2$ and $M=0.10\,m$.  
The average momentum, $\vec{\bar{P}}$, is considered up to $50\,m$,  
which includes the Lab frame ($\vec{\bar{P}} \simeq 25\,m$). Notice  
that the example is a somewhat extreme one. The large drop off of the  
single-particle contribution as well as the large contribution of the  
two-body currents to restore the covariance has more to do with the  
low mass of the system than with the $Q^2$ value which is not excessively  
high. This example has been given to illustrate the role of two-body  
currents in a particular case where the various contributions can be  
dealt with easily. These ones are part of more realistic calculations  
with finite-range forces, for which we expect further contributions  
however.  

A study similar to the above one could be performed in the front-form approach 
for the configuration,  $ \vec{Q} \parallel (\vec{P}_i+\vec{P}_f) \parallel 
\vec{n}$. The results for the contribution of the single-particle current can 
significantly differ from those where  $ \vec{Q} \perp (\vec{P}_i+\vec{P}_f)$, 
generally considered as the most reliable. These ones are recovered when one 
adds the contribution of two-body currents that have the same origin as those 
given by Eq. (\ref{40b}) for the instant-form case.
 
\noindent 
$\bullet$ For the transverse configuration, where the energies of the  
initial and final states remain equal ($E_i = E_f$), we did not find  
much sensitivity. The departure is of the order of what the omission  
of possible factors such as $c.f.(E_i=E_f)$ given by Eq.~(\ref{30a3}),  
will produce. It also compares to  
the difference between the instant-form and the front-form results  
presented in Tables~\ref{t10}-\ref{t30} (roughly 10\% and 30\% at most  
for the form factors $F_1$ and $F_0$, respectively). As expected, the  
instant-form results tend to the front-form ones when the average  
momentum, $\vec{\bar{P}}$, goes to $\infty$. The small sensitivity  
of the results on  $\vec{\bar{P}}$  is evidently related to the  
orthogonality  property of this vector with the momentum transfer,   
$\vec{Q}$, which minimizes the interference effects that plague  
the results for the parallel configuration. Interestingly, the  
introduction of the correcting factor given by Eq.~(\ref{30a3}) would  
remove a sizable part of the difference between the instant- and  
front-form calculations of $F_1$ and even the totality for the infinite  
boson-mass case.

\subsection{Improved interaction} 
We noticed that form factors in the instant and front forms for the zero-mass  
boson case evidence a departure from the ``exact'' ones (B.S.) by a  
factor two or so at the highest $Q^2$ values. For the main purpose of  
this paper, which is to look at the gross features evidenced by  
different approaches, this is rather sufficient, knowing that one of  
the approaches shows orders of magnitude discrepancies. For the purpose  
of better checking the theoretical framework, one can nevertheless  
wonder about this slight discrepancy and its relation to the choice  
of the interaction which was taken as the simplest one possible.  
No attempt was made for reproducing the spectrum of the ``exact''  
one, which is known to be close to the Coulombian one with an effective  
strength~\cite{Amghar:2000pc}. As departures from this spectrum are due to the  
normalization factors, $1/\sqrt{e_k}$ in Eq.~(\ref{20c}), which are  
essential on the other hand, a simple way to partially correct for  
their effect is to compensate for the $k^2/m^2$ dependence they  
produce at the lowest order in this quantity. While doing so, the  
correction should not change the high momentum behavior of the   
interaction so that expectations from the consideration of the Born  
amplitude are still fulfilled. A better interaction could thus be: 
\begin{equation} 
V=-g^2\, 
\frac{4\,m^2}{\mu^2+(\vec{k}-\vec{k}\,')^2} \; 
\left(1+\frac{e_k-m}{2\,e_k}+\frac{e_{k'}-m}{2\,e_{k'}}\right). 
 \label{40m} 
\end{equation} 
For small $k$, this interaction, including the normalization factors,  
looks more like a Coulombian one. We also expect its strength in this  
range to be closer to the latter one given in Table ~\ref{t40}, what  
we indeed find. Interestingly, the force at high $k$ is enhanced by  
a factor 2, partly compensating the effect of the renormalization of the  
interaction due to retardation effects and becoming closer to the bare  
one. Although this result  is conceivable, we have not checked whether  
the observation is fully relevant. This could provide a useful constraint   
on the derivation of the effective interaction. 
 
\begin{table}[tb!] 
\caption{Elastic scalar- and vector-  form factors, $F_1(Q^2)$ and  $F_0(Q^2)$,  
with a tentatively improved interaction:  
Same as in Table~\protect\ref{t10}, but  with wave functions obtained from the  
interaction model, Eq.~(\protect\ref{40m} ).  
The coupling constants corresponding to the two masses $M=1.568\,m$ and  
$M=0.10\,m$ are, respectively,  $g^2/4\pi=1.314$ and  
$2.267$. These ones can be compared to the ones for the Wick-Cutkosky model,     
$g^2/4\pi= 3$ and $g^2/4\pi=1.996\,\pi$ and those for the Coulombian potential 
used  in the non-relativistic case, $g^2/4\pi=1.241$ and $1.997$, to which 
they are  becoming closer. Results for  the Wick-Cutkosky model are recalled. 
\label{t50} } 
\medskip 
\begin{center} 
\begin{tabular}{lcccccc} 
\hline  \rule[0pt]{0pt}{3ex} 
  $Q^2/m^2$             &   0.01   &   0.1  &  1.0   &  10.0 & 100.0 
  \\ [1.ex] \hline 
 $M=1.568\,m$                       &          &        &        &       &   \\ 
 $F_1\;\;\;\;$I.F.  & 0.997  & 0.965 & 0.723 & 0.159 & 0.66-02  \\ [0.ex] 
 $F_0\;\;\;\;$I.F.  & 1.247  & 1.201 & 0.860 & 0.153 & 0.48-02  \\ [0.ex] 
 $F_1\;\;\;\;$F.F.  & 0.996  & 0.963 & 0.712  & 0.147  & 0.58-02 \\ [0.ex] 
 $F_0\;\;\;\;$F.F.  & 1.079  & 1.039 & 0.742  & 0.132  & 0.43-02 \\ [0.ex] 
 $F_1\;\;\;\;$P.F. & 0.994  & 0.942  & 0.576  & 0.329-01 & 0.37-04 \\ [0.ex] 
 $F_0\;\;\;\;$P.F. & 0.993  & 0.933  & 0.527  & 0.219-01 & 0.19-04 \\ [1.ex] 
 $F_1\;\;\;\;$B.S.   & 0.996   & 0.962  & 0.705  &  0.139  & 0.50-02  \\ [0.ex]  
 $F_0\;\;\;\;$B.S.   & 1.123   & 1.080  & 0.767  &  0.132  & 0.39-02 \\ [1.ex]  
 \hline 
 $M=0.1\,m$                       &          &        &        &       &   \\ 
 $F_1\;\;\;\;$I.F.  & 0.999 & 0.985 & 0.862 & 0.357 & 0.285-01  \\ [0.ex] 
 $F_0\;\;\;\;$I.F.  & 1.496 & 1.465 & 1.205 & 0.375 & 0.211-01 \\ [0.ex] 
 $F_1\;\;\;\;$F.F.    & 0.998  & 0.983  & 0.847 &  0.332 & 0.260-01 \\ [0.ex]  
 $F_0\;\;\;\;$F.F.    & 1.153  & 1.130  & 0.937 &  0.306 & 0.187-01 \\ [0.ex]  
 $F_1\;\;\;\;$P.F. &0.517 & 0.149-01 & 0.91-05 & 0.170-08 & 0.24-12 \\ [0.ex] 
 $F_0\;\;\;\;$P.F. &0.431 & 0.087-01 & 0.46-05 & 0.085-08 & 0.12-12 \\ [1.ex] 
 $F_1\;\;\;\;$B.S.   & 0.998  & 0.983   & 0.848  &  0.338  & 0.283-01 \\ [0.ex]  
 $F_0\;\;\;\;$B.S.   & 1.247  & 1.222   & 1.016  &  0.338  & 0.217-01 \\ [1.ex]  
\hline \\ 
\end{tabular} 
\end{center} 
\end{table} 
 
The form factors calculated with the above improved interaction are given  
in Table~\ref{t50} while  the spectrum of the model for the ground state,  
its first radial excitation and the $l=1$ state are given in   
Table~\ref{t40}. It is seen that the form factors become closer to or  
slightly overshoot the ``exact'' ones. At the same time, when it is  
compared to the simplest choice of the interaction, the spectrum 
with the improved interaction becomes better. The first radially-excited  
and the first $l=1$ states tend to be closer to each other and  
closer to the ``exact'' ones.  
 
The aim of the present subsection was to provide some insight on a
possible  sensitivity of form factors to the effective interaction used  
in the calculation of wave functions. This was done by modifying the  
initial interaction in a manner that would be more consistent with the  
main features of the Wick-Cutkosky model, but we have not  
attempted a refined description. 
Considering the simultaneous improvement of the form factors at 
the highest $Q^2$ values and the position of the first radially excited 
state, we expect that such a refined interaction could easily remove a large 
part of the remaining differences. This should be checked carefully however,  
especially in view of doubts sometimes expressed about reproducing in  
relativistic quantum mechanics approaches results based on the Bethe-Salpeter 
equation.

From the present study, and taking into account the constraints that we 
imposed  on the single-particle currents, it thus appears that Breit-frame 
instant-form  and front-form approaches based on these currents provide 
form factors close to  each other as well as close to the ``exact'' ones. 
The role of two-body currents becomes essential when considering kinematical
configurations with ($E_i \neq E_f$) for the instant form and ($P^+_i \neq 
P^+_f$) for the front form.

\section{Discussion and conclusion} 
In this work, we have studied the sensitivity of form factors of a  
two-body system to various ways of implementing boost effects.
The emphasis has been put on the instant form of relativistic quantum 
mechanics but, as is well known, the corresponding mass operator can be 
employed for front and point forms of this framework. Since we wanted 
to make some relation with a field-theory model, we proceeded 
in a way that perhaps differs from other ones. We nevertheless 
recovered all the ingredients that allowed Bakamjian and Thomas 
to construct the generators of the Poincar\'{e} group. 
The system under consideration consists 
of two scalar particles interacting by the exchange of another scalar  
one. Only the ground state has been considered. We looked at different  
values of the exchanged boson mass, $\mu/m= 0,\; 0.15 ,\; \infty$.  
In the first and last cases, results that can be considered as ``exact''  
ones are presented. Results obtained by applying a Galilean boost are  
also given. Dealing with degrees of freedom that have necessarily an effective 
character, there are uncertainties. The first one has to do with the one- and 
the two-body parts of the current in relativistic  quantum mechanics approaches. 
In all cases, we required that the same one-body current be recovered in the 
small coupling limit. Beyond, for strongly bound systems, it appeared that the 
electromagnetic current should incorporate improvements to remove an undesirable 
feature relative to the ratio of the scalar and charge form factors. This 
correction, which is part of the developments that  should be accounted for in 
the future, has been done in the same way for the instant- and point-form 
approaches (the standard front-form results are unaffected by this change). It 
offers other advantages that let us think that it is a necessary ingredient in 
the description of the current. Another correction has been introduced for the 
scalar current in the instant form but its role is a minor one.  
The second uncertainty has to do with the effective interaction when an 
``exact'' calculation is available. This concerns the zero-mass case, the 
infinite mass case being insensitive to the interaction itself 
once the mass of the system is  given. 
 
Within the above uncertainties, most approaches agree with each other  
and sometimes produce the same results. In the range of $Q^2$ considered  
here, $0 \leq Q^2 \leq 100\, m^2$, the discrepancy between the instant-  
or front-form results and the ``exact'' ones does not exceed a factor 2.  
From earlier studies, it is reasonable to attribute this discrepancy to  
the interaction model. This one is too simple to account for the full  
mass spectrum of the ``exact'' calculation. Amazingly, a non-relativistic  
approach, mainly characterized by a Galilean boost, is not bad in a  
regime of large binding energies or large $Q^2$, which is highly  
relativistic. Only the form factors in the point form depart from the other  
results, confirming what has been obtained elsewhere with different  
inputs~\cite{Desplanques:2001zw,Desplanques:2001ze}.  
Two main features should be emphasized: 
\begin{itemize} 
\item The form factors at high $Q^2$ have most often the wrong power  
law behavior, missing the expected Born amplitude.  
\item The charge or scalar radii tend to infinity when the total  
mass of the system tends to zero while all the other approaches lead  
to a finite value.  
\end{itemize} 
It is noticed that the choice of the form of the single-particle current  
has little influence on the discrepancy (a factor 2 at most) as 
if minimal (essentially kinematical) consistency requirements 
in the different approaches were fixing this part. The discrepancy 
therefore points out to the way the boost is incorporated 
in the wave functions, which is more sensitive to the dynamics. 
In this respect, one notices that the  
corresponding expressions entering the form factors involve at many places  
the product of the momentum transfer, $Q$, and a quantity $2e_p/M$.  
This last factor, which is larger than one, leads to an enhanced  
effective momentum transfer, largely explaining the two features  
emphasized above. It is absent in all the other  
approaches~\cite{Desplanques:2001ze}. 

Despite uncertainties, there is a cumulative evidence that the form factors  
in the point form differ  from all the other ones when calculated 
from a single particle current. Evidently, this problem can be solved  by 
including two-body currents~\cite{Desplanques:2001}.  
These ones should  
also be present in the other approaches but it is clear that their role  
in the point-form calculations is more essential. In this case, one has to  
correct for orders of magnitude while, in the other ones, one expects   
contributions of the same size as the single-particle current. In  
view of such a large discrepancy, one can however wonder whether the  
implementation of the point-form approach for the calculation of the  
single-particle current itself is the most convenient one. 
 
In this respect, it is reminded that the total momentum operator for a two-body 
system in the point-form approach may be written as:
\begin{equation}
P^{\mu}=p_1^{\mu}+p_2^{\mu}+P^{\mu}_I,
\end{equation}
where $P^{\mu}_I$ represents the interaction part. In the implementation of the 
point-form used until now, this  term assumes a form proportional to the  
4-velocity of the system ($\lambda^{\mu}$ in our notations). This is the  
simplest choice one can think of. It is not unique however and one can 
therefore  wonder whether other choices could minimize the problem raised by 
present results.

In the original work by Dirac, the point-form approach is described on  
a hyperbolo\"{\i}d, $x \cdot x=ct$. Some studies along these lines have  
been made~\cite{Fubini:1973mf,Gromes:1974yu} but they did not go quite far, 
due perhaps  
to the related nonlinear constraint between the coordinates, $x^0$ and  
$\vec{x}$. In practice, recent applications of the point-form  
approach~\cite{Desplanques:2001zw,Desplanques:2001ze,Allen:2000ge,Wagenbrunn:2000es,Coester:1998ih,Coester:1998ki}  
rely on employing wave functions  
issued from a mass operator whose solutions can also be identified 
with instant-form ones in the center of mass system. These last solutions 
are characterized by the description of the dynamics on a hyper-plane,  
$\lambda_0 \cdot x=ct$, with $\lambda_0^{\mu}=(1,0,0,0)$. As noticed 
by Sokolov \cite{Sokolov:1985jv}, this point-form approach is not 
identical to the one proposed by Dirac. While the surface,  
$x \cdot x=ct$, is invariant under a Lorentz transformation, the  
surface, $\lambda_0 \cdot x=ct$, is not. It is our opinion that the  
replacement of the former surface by the latter, which may be roughly  
appropriate for an isolated system, is not adequate for  
a process like elastic scattering, which requires two systems with  
different momenta, $P_i^{\mu}$ and  $P_f^{\mu}$.  
 
Indeed, when the system at rest described on the hyper-plane, $\lambda_0  
\cdot x=ct$, is kinematically boosted to get initial and final states  
with four-momenta, $P_i^{\mu}$ and  $P_f^{\mu}$, these ones appear  
as described (quantized) on different surfaces,  $\lambda_i \cdot x=ct$  
and  $\lambda_f \cdot x=ct$, where $\lambda_{i,f}^{\mu} \propto  
P_{i,f}^{\mu}$ with $\lambda^2=1$. This feature results 
from the identification of point- and instant-form wave functions 
in the center of mass. It does not correspond to  the usual description 
of a process which, generally, relies on the same definition  
of the surface at all steps. As a measure of the difficulties,  
it is remarked that time has not the same meaning for an observer  
related to the initial or the final state, preventing one from defining  
time-ordered diagrams. Actually, a large part of the difficulty for  
implementing relativity in quantum mechanics approaches is precisely  
to determine how to appropriately boost a system while keeping this  
constraint. It is also noticed that a system at rest could  
be described as well on hyper-planes, $\lambda \cdot x=ct$, with  
$\lambda^{\mu} \neq \lambda_0^{\mu}$~\cite{Szczepaniak:1995mi} and there 
is no  compelling argument for choosing the particular hyper-plane,  
$\lambda_0 \cdot x=ct$, except perhaps for simplicity. As is well  
known, changing the hyper-plane implies the dynamics but the problem  
is not much more complicated than determining the relation between  
wave functions for different $\vec{P}$ in the instant-form approach,  
which we presented in Sec. 2. 

As a complementary remark, we would 
like to stress that the kinematical boost retained in the point-form 
approach appears practically as an approximation to the 
instant-form boost. The latter one, as can be seen in Eq.~(\ref{20e3}), also 
contains an interaction-dependent term whose role is essential 
for the derivation of an invariant mass operator within the instant-form 
approach, as well as for reproducing the exact form factors in some cases. 
Examining the transformation, it is noticed that this second term depends 
on the total momentum, preventing one to split the instant-form boost 
operation into independent kinematical and dynamical parts.  In the 
point-form approach however, there is the possibility that this was feasible. 
One could  imagine for instance that the wave function in terms 
of the physical momenta of the constituent particles, $\vec{p}_1$ 
and $\vec{p}_2$, is related to the center of mass one by a kinematical 
boost but that the relation of this last one to the solution of the mass  
operator, given in terms of the total momentum $\vec{P}$ and the internal 
variable  $\vec{k}$, involves the dynamics. 
With this respect, it is noticed that
the condition for the momenta, $\vec{p}_1+\vec{p}_2=0$, is consistent with 
the instant form, due to the kinematical character of the momentum,
but there is no reason that this holds in other forms where the momentum has, 
partly or totally, a dynamical character, as can be seen in Eq.~(\ref{20ba}). 
The introduction of the dynamics as suggested above would greatly contribute 
to reduce the qualitative gap between the point-form approach used here,  
where the boost transformation is insensitive to the dynamics, 
and the other forms. This would also 
be more consistent with the existence of a unitary transformation 
between the different forms \cite{Sokolov:1977im}.
 
While some conditions for applying the point-form approach, such as 
the  independence of the interaction on the velocity, have been 
mentioned in the  literature, it sounds that the above aspect relative 
to the nature of the  surface, on which the system is described in 
practice, has been overlooked. The problem concerns less the 
construction of the generators of the Poincar\'{e} algebra in 
terms of the total momentum $\vec{P}$ and the internal variable than 
the relation of this set of variables to the physical ones. In  view 
of the previous discussion, we believe that some improvement can be  
introduced, at the expense of making calculations more complicated than 
they  were until now, but not much more than some of the instant-form  
calculations presented in this work. The idea is to start with states at rest  
described on two different surfaces, $\lambda_i' \cdot x=ct$ and $\lambda_f' 
\cdot x=ct$, in such a way that these ones coincide with a unique one,  
$\lambda \cdot x=ct$, after a boost is performed for getting states with 
the  appropriate momenta. Providing a better account of the relative 
boost of initial  and final states, this should be sufficient to remove  
the main differences between  the point-form results and the other ones,  
which appear at small $M$ or large $Q^2$ when only the contribution of the 
single-particle current is considered. 
In  the spirit of the point-form approach possibly described on a 
hyperbolo\"{\i}d, one can go a step further and 
assume that  the four vector, $\lambda^{\mu}$, is not a fixed one, like 
in the instant-form  approach, but is allowed to change like any four-vector 
so that the surface,  $\lambda \cdot x=ct$, is unchanged when a Lorentz  
transformation is made. In this order, the four-vector $\lambda^{\mu}$ 
has to be  chosen as a combination of the four vectors entering the process. 
An immediate  choice, which is consistent with the symmetry of the initial 
and final states  for elastic scattering, is: 
\begin{equation} 
\lambda^{\mu}= \frac{P_f^{\mu}+P_i^{\mu}}{\sqrt{(P_f+P_i)^2}}. 
\label{50a} 
\end{equation} 
This choice also guarantees that the divergence of the matrix element of the  
electromagnetic current is zero.  However, it does not imply that the current  
itself is conserved as it should be. This generally requires considering the  
contribution of two-body currents. The above choice also allows one  
to recover the prescription made in previous works for an isolated system,  
since in this case  $P_i^{\mu}=P_f^{\mu} \propto \lambda^{\mu}$, which is  
nothing but the four velocity introduced there.  
 
It is straightforward to show that a calculation of form factors as described  
above would provide results identical to those we obtained in the Breit frame  
in instant-form, as presented in Tables \ref{t10}-\ref{t30}. 
As is seen from these  
tables, this improved implementation of the point-form approach now  
compares with the other approaches. While starting from the instant-form, 
the corresponding results can be made covariant via the  
replacement of the four-vector, $\lambda_0^{\mu}$, by $\lambda^{\mu}$,  
as defined in Eq.~(\ref{50a}), in all steps of the calculation. 
The identity with the  
form factors in the instant form  only holds in the Breit frame, however. In  
absence of two-body currents restoring the covariance, these form factors  
in the instant form will generally be different in another frame.  
 
This improved implementation  of the point-form approach, like recent ones,  
is  an approximation to the one originally proposed by Dirac. It has some  
of its characteristics but it has certainly limitations that have to be  
explored. The fact that one recovers results obtained by  
other approaches is an indication that we are on the good track.  
On the other hand, different relativistic-quantum  mechanics approaches 
should be equivalent up to a unitary transformation. This was  
hardly conceivable with the earlier implementation of the point-form approach, 
where initial and final states appeared as described on different hyper-planes.  
This difficulty is removed with the approach presented above. 
 
Present results for various forms of relativistic quantum mechanics suggest 
that, in our model, there is a problem with the implementation of 
the point-form one.  
In this last case, either large contributions of two-body currents are 
required or the present implementation of the point-form approach 
for calculating the contribution of the single-particle 
current has to be significantly improved. This conclusion 
is free of uncertainties due to spin complications or to  
intrinsic form factors of the constituents, for instance. 
Under these conditions,  
one may wonder about the agreement with experiment obtained  
in~\cite{Wagenbrunn:2000es} of the nucleon form factors. 
We notice that the main effect encountered there has the same  
origin as in the present case. It arises for a large part from an effective 
enhancement  
of the momentum transfer by a factor $\sum e_k/M$, representing the ratio  
of the kinetic energies to the total mass. If this  
effect is the result of an incomplete account of interaction effects, as  
we believe, other explanations for the nucleon form factors have to 
be found. Actually, it is known for a long time that nucleon form factors  
in the lowest $Q^2$ range can be largely accounted for by the vector-meson  
dominance phenomenology (see ref.~\cite{Cano:2001hy} for an attempt of a  
microscopic account of this physics). At high $Q^2$, there are indications  
that the fall off of the magnetic nucleon form factors is too fast as compared  
to the measurements, quite in agreement with our findings for scalar-particle 
systems. This points to another fault of these results. In comparison  
to a non-relativistic calculation, a proper account of relativity is expected  
to enhance the form factors at high $Q^2$: a 
non-relativistic approach would predict a  $Q^{-8}$ fall off   
($Q^{-6}$ with semi-relativistic kinetic energy) while the QCD  
expectation is $Q^{-4}$. The difference is, in first approximation, due to 
factors $(e+m)/(2\,m)$ that are 
replaced by $1$ in the non-relativistic case~\cite{Desplanques:2000ev}.  
The point-form results, as presented here and in other works, 
go in the other direction.  
 
It was found that binding energies calculated from the Bethe-Salpeter  
equation in ladder approximation could be understood in a  
quantum-mechanics scheme with a theoretically motivated, effective  
interaction going beyond the standard instantaneous 
approximation~\cite{Amghar:2000pc}.
Present results show that form factors calculated from solutions  
of this equation can be reasonably understood within relativistic  
quantum mechanics (we assume that the problem for the point-form 
approach will be solved according to our proposal or equivalent ones).
These results strongly suggest that the two approaches can lead to the same 
predictions,
a principle which is not universally  
accepted. Taking this principle for granted, many improvements can be  
considered and tested. They concern the effective interaction and the  
structure of the current, in particular in relation with its two-body part. 
A next step could include the introduction of spin, required  
for the study of realistic systems, or the extension to three-body 
systems. For the time being, and despite its partly academic character,  
we believe that the present work can be useful for improving the  
implementation of relativity in the description of few-body systems,  
as for the Lorentz-contraction or the asymptotic behavior of  
form factors, for instance.

{\bf Acknowledgments} 
The authors are very grateful to S. Noguera for stimulating discussions  
regarding the role of the hyperbolo\"{\i}d surface in the point-form approach.  
We acknowledge V. Sauli for providing some results concerning the  
calculation of form factors in the case of a finite boson mass, using solutions  
of the Bethe-Salpeter equation. 
This work has been supported by the 
EC-IHP Network ESOP, under contract HPRN-CT-2000-00130.
 

\begin{thebibliography}{10} 
\expandafter\ifx\csname url\endcsname\relax 
  \def\url#1{\texttt{#1}}\fi 
\expandafter\ifx\csname urlprefix\endcsname\relax\def\urlprefix{URL }\fi 
 
\bibitem{Zuilhof:1980ae} 
M.~J. Zuilhof, J.~A. Tjon, Phys. Rev. C 22 (1980) 2369. 
 
\bibitem{Gross:1983yt} 
F.~Gross, B.~D. Keister, Phys. Rev. C 28 (1983) 823. 
 
\bibitem{Karmanov:1992fv} 
V.~A. Karmanov, A.~V. Smirnov, Nucl. Phys. A 546 (1992) 691. 
 
\bibitem{Carbonell:1998rj} 
J.~Carbonell, B.~Desplanques, V.~A. Karmanov, J.~F. Mathiot,  
Phys. Rept. 300 (1998) 215. 
 
\bibitem{Licht:1970pe} 
A.~L. Licht, A.~Pagnamenta, Phys. Rev. D 2 (1970) 1150. 
 
\bibitem{Friar:1973} 
J.~L. Friar, Ann. Phys. 81 (1973) 332. 
 
\bibitem{Chung:1988my} 
P.~L. Chung, F.~Coester, B.~D. Keister, W.~N. Polyzou, Phys. Rev. C 37 (1988) 
  2000. 
 
\bibitem{Chung:1988mu} 
P.~L. Chung, F.~Coester, W.~N. Polyzou, Phys. Lett. B 205 (1988) 545. 
 
\bibitem{Lev:1995} 
F.~Lev, Ann. Phys. 237 (1995) 355. 
 
\bibitem{Klink:1998} 
W.~H. Klink, Phys. Rev. C 58 (1998) 3587. 
 
\bibitem{Desplanques:2001zw} 
B.~Desplanques, L.~Theu{\ss}l, Eur. Phys. J. A. 13 (2002) 461.
 
\bibitem{Desplanques:2001ze} 
B.~Desplanques, L.~Theu{\ss}l, S.~Noguera,  Phys. Rev. C 65 (2002) 038202. 
 
\bibitem{Salpeter:1951sz} 
E.~E. Salpeter, H.~A. Bethe, Phys. Rev. 84 (1951) 1232. 
 
\bibitem{Wick:1954eu} 
G.~C. Wick, Phys. Rev. 96 (1954) 1124. 
 
\bibitem{Cutkosky:1954ru} 
R.~E. Cutkosky, Phys. Rev. 96 (1954) 1135.

\bibitem{Acus:2000ah}
A.~Acus, E.~Norvaisas, D.O.~Riska, Physica Scripta 64 (2001) 113.

\bibitem{Santopinto:2002}
E. Santopinto, invited talk presented at BARYON2002, 
Newport News (March 3-8, 2002). 
 
\bibitem{Lepage:1979zb} 
G.~P. Lepage, S.~J. Brodsky, Phys. Lett. B 87 (1979) 359. 
 
\bibitem{Lepage:1979za} 
G.~P. Lepage, S.~J. Brodsky, Phys. Rev. Lett. 43 (1979) 545; ibid. 1625. 
 
\bibitem{Alabiso:1974sg} 
C.~Alabiso, G.~Schierholz, Phys. Rev. D 10 (1974) 960. 
 
\bibitem{Hamme:1992} 
B.~Hamme, W.~Gl{\"o}ckle, Few Body Syst. 13 (1992) 1. 

\bibitem{Glockle:1987hb} 
W.~Gl{\"o}ckle, Y.~Nogami, Phys. Rev. D 35 (1987) 3840. 
 
\bibitem{Allen:2000ge} 
T.~W.~Allen, W.~H. Klink, W.~N. Polyzou, Phys. Rev. C 63 (2001) 034002. 
 
\bibitem{Wagenbrunn:2000es} 
R.~F.~Wagenbrunn, S.~Boffi, W.~Klink, W.~Plessas, M.~Radici,  
  Phys. Lett. B 511 (2001) 33. 
 
\bibitem{Cano:2001hy} 
F.~Cano, B.~Desplanques, P.~Gonzalez, S.~Noguera, 
Phys. Lett. B 521   (2001) 225. 
 
\bibitem{Bakamjian:1953kh} 
B.~Bakamjian, L.H.~Thomas, Phys. Rev. 92 (1953) 1300.

\bibitem{Keister:1991sb} 
B.~Keister, W.~Polyzou, Adv. Nucl. Phys. 20 (1991) 225.

\bibitem{Coester:1975hj} 
F.~Coester, A. Ostebee, Phys. Rev. C 11 (1975) 1836.

\bibitem{Sokolov:1985jv} 
S.~N. Sokolov, Theor. Math. Phys. 62 (1985) 140.

\bibitem{Wallace:2001nv} 
S.~J. Wallace, Phys. Rev. Lett. 87 (2001) 180401. 
 
\bibitem{Gross:1966fg}
F. Gross, Phys. Rev. 142 (1966) 1025.

\bibitem{Lev:1994vf}
F. Lev, Phys. Rev. D 49 (1994) 383.

\bibitem{Sokolov:1977im} 
S.~N. Sokolov, A.~N. Shatnii, Theor. Math. Phys. 37 (1978) 1029.

\bibitem{Amghar:2000pc} 
A.~Amghar, B.~Desplanques, L.~Theu{\ss}l, Nucl. Phys. A 694 (2001) 439. 
 
\bibitem{Fukuda:1954} 
N.~Fukuda, K.~Sawada, M.~Takateni, Prog. Theor. Phys. 12 (1954) 156. 
 
\bibitem{Okubo:1954} 
S.~{\^O}kubo, Prog. Theor. Phys. 12 (1954) 603; 
M.~Sugawara, S.~{\^O}kubo, Phys. Rev. 117 (1960) 605. 
 
\bibitem{Desplanques:2001} 
B.~Desplanques, L.~Theu{\ss}l, in preparation. 
 
\bibitem{Sauli:2001zz} 
V.~Sauli, \texttt{hep-ph/0111433}. 
 
\bibitem{Sauli:2001zy} 
V.~Sauli, J.~Adam, Jr., \texttt{hep-ph/0110090}. 

\bibitem{Mandelstam:1955sd}
S. Mandelstam, Proc. Roy. Soc. London A 233 (1955) 248.
 
\bibitem{Desplanques:2000} 
B.~Desplanques, R.~Medrano, S.~Noguera,  L.~Theu{\ss}l,  
Contribution to the 17th European Conference on Few-Body Problems in Physics,  
11-16 September 2000, \'Evora. 
 
\bibitem{Desplanques:2000ev} 
B.~Desplanques, B.~Silvestre-Brac, F.~Cano, P.~Gonzalez, S.~Noguera,  
Few Body Syst. 29 (2000) 169. 
 
\bibitem{Szczepaniak:1995mi} 
A.~Szczepaniak, C.~R. Ji, S.~R. Cotanch, Phys. Rev. C 52 (1995) 2738. 
 
\bibitem{Fubini:1973mf} 
S.~Fubini, A.~J. Hanson, R.~Jackiw, Phys. Rev. D 7 (1973) 1732. 
 
\bibitem{Gromes:1974yu} 
D.~Gromes, H.~J. Rothe, B.~Stech, Nucl. Phys. B 75 (1974) 313. 
 
\bibitem{Coester:1998ih} 
F.~Coester, D.~O. Riska, Few Body Syst. 25 (1998) 29. 
  
\bibitem{Coester:1998ki} 
F.~Coester, K.~Dannbom, D.~O. Riska, Nucl. Phys. A 634 (1998) 335. 
 
\end{thebibliography}

\end{document}